\documentclass[pra,aps,showpacs,twocolumn,amsmath,amssymb,floatfix]{revtex4}
\usepackage{graphicx}
\usepackage{amssymb,amsmath}
\usepackage{subfigure}
\usepackage{here}
\usepackage{bbm}
\usepackage[dvips]{color}

\newcommand{\surfacePlotScale}[0]{0.21}

\newcommand{\Rb}{$^{87}$Rb}

\newcommand{\hc}{\mathrm{h.c.}}

\newcommand{\tr}{\mathrm{Tr}}

\newcommand{\Hhat}{\hat{H}}
\newcommand{\bhat}{\hat{b}}
\newcommand{\bhatd}{\hat{b}^\dagger}
\newcommand{\nhat}{\hat{n}}
\newcommand{\Ihat}{\hat{\mathbbm{1}}}
\newcommand{\rhohat}{\hat{\rho}}
\newcommand{\thetahat}{\hat{\theta}}

\newcommand{\schrod}{Schr\"{o}dinger~}

\begin{document}

\title{Quantum Many-Body Dynamics of Dark Solitons in Optical Lattices}
\author{R. V. Mishmash$^{1,2,3}$, I. Danshita$^{3,4,5}$, Charles W.
Clark$^3$, and L. D. Carr$^{2,3}$}
\affiliation{
$^1$Department of Physics, University of California, Santa Barbara, CA 93106, USA \\
$^2$Department of Physics, Colorado School of Mines, Golden, CO 80401, USA \\
$^3$Joint Quantum Institute, National Institute of Standards and Technology, Gaithersburg, MD 20899, USA \\
$^4$Department of Physics, Boston University, Boston, MA 02215, USA \\
$^5$Department of Physics, Faculty of Science, Tokyo University of Science, Shinjuku-ku, Tokyo 162-8601, Japan
}
\date{\today}

\begin{abstract}
We present a fully quantum many-body treatment of dark solitons formed by ultracold bosonic atoms in one-dimensional optical lattices.  Using time-evolving block decimation to simulate the single-band Bose-Hubbard Hamiltonian, we consider the quantum dynamics of density and phase engineered dark solitons as well as the quantum evolution of mean-field dark solitons injected into the quantum model.  The former approach directly models how one may create quantum entangled dark solitons in experiment.  While we have already presented results regarding the latter approach elsewhere [Phys. Rev. Lett. {\bf 103}, 140403 (2009)], we expand upon those results in this work. In both cases, quantum fluctuations cause the dark soliton to fill in and may induce an inelasticity in soliton-soliton collisions.  Comparisons are made to the Bogoliubov theory which predicts depletion into an anomalous mode that fills in the soliton.  Our many-body treatment allows us to go beyond the Bogoliubov approximation and calculate explicitly  the dynamics of the system's natural orbitals.
\end{abstract}

\pacs{03.75.-b, 03.75.Gg, 03.75.Lm, 05.45.Yv}

\maketitle

\section{Introduction} \label{sec:Intro}

Solitons are robust nonlinear waves that appear in many facets of nature.  They are localized, persistent in time, and collide elastically.  Solitons with both phase and amplitude are often described to lowest order in systems with weak local nonlinearity by the famous nonlinear Schr\"odinger equation (NLS)~\cite{Kivshar98_PhysRep_298_81}.  For example, the NLS describes twist and curvature of a string in three dimensions~\cite{lamb1976}.  However, corrections beyond the NLS, necessary in order to more closely model soliton properties, are less universal.  For instance, for solitons in optical fibers one must consider higher order dispersion.  Bose-Einstein condensates (BECs) are also well-known to be modeled by the NLS, or Gross-Pitaevskii equation (GP).  Dark solitons were theoretically predicted in BECs~\cite{reinhardt1997,kevrekidisPG2008} and subsequently have been observed in a number of experiments~\cite{Burger99_PRL_83_5198,Denschlag00_Science_287_97,anderson2001,ginsberg2005,Sengstock08_NaturePhys_4_496,Sengstock08_PRL_101_120406, Oberthaler08_PRL_101_130401}.  The NLS then represents taking the lowest order average of quantum operators from the quantum many-body description; therefore, it is called a \emph{mean-field theory}.  One naturally expects thermal or quantum corrections to mean-field theory. Regarding thermal corrections, recent finite temperature studies~\cite{Jackson07_PRA_75_051601} based on coupling Boltzmann equations to the NLS have indicated that thermal fluctuations cause dark solitons to decay. The question then arises, are quantum corrections, called quantum fluctuations, sufficiently large so as to change the basic properties of dark solitons, a fundamental prediction of mean-field theory and a ubiquitous property of weakly nonlinear systems?

The answer so far has been a resounding no.  First, BEC soliton experiments have worked in regimes with very small quantum fluctuations.  Such fluctuations can be calculated quantitatively with the Bogoliubov-de-Gennes equations (BDG) for weakly interacting BECs, and one finds much less than 1\% for the system's ground state~\cite{Dalfovo99_RevModPhys_71_463}. Second, BDG methods have been used to describe quantum fluctuations of solitons.  The main result is that within BDG framework the dark soliton, which manifests as a density notch with a phase jump across it, appears to be filled in:  the soliton's position is uncertain, so that in an ensemble average of measurements it appears blurred ~\cite{Dziarmaga03_JPhysB_36_1217, Dziarmaga06_JPhysB_39_57, Dziarmaga04_PRA_70_063616}.  However, these studies predict that in any particular measurement the soliton is indeed localized, and the predictions of mean-field theory for the density, phase, and stability properties of the soliton hold.

In this article, we provide new answers to the question of how quantum fluctuations affect dark solitons, expanding on the results which two have us have presented in Ref. \cite{Mishmash09_short}.  We show that in a regime of stronger quantum fluctuations, the results based on BDG theory no longer hold.  Solitons are not just delocalized in an ensemble measurement, but do in fact decay in the zero temperature quantum setting.  They even collide inelastically.  By use of a recently invented method for entangled quantum many-body dynamics in one dimension~\cite{Vidal04_PRL_93_040502}, we show explicitly how both mean-field theory and BDG theory break down.  To obtain large quantum fluctuations we make use of an optical lattice~\cite{Lewenstein07_AdvPhys_56_243, Bloch08_RevModPhys_80_885}.  BECs in optical lattices are known to have two distinct quantum phases, the Mott insulator and the superfluid \cite{Fisher89_PRB_40_546}.  We emphasize that throughout our study we remain in the superfluid phase.  We also treat higher order correlations, a key signal for dark soliton decay in a fully quantum theory, where the mean field does not force symmetry breaking on the system~\cite{Carr08_PRL_100_060401,Carr09,Carr09b}.

\begin{figure}[t]
\begin{center}
\hspace{-0.2in}\subfigure{\scalebox{0.27}{\includegraphics[angle=0]{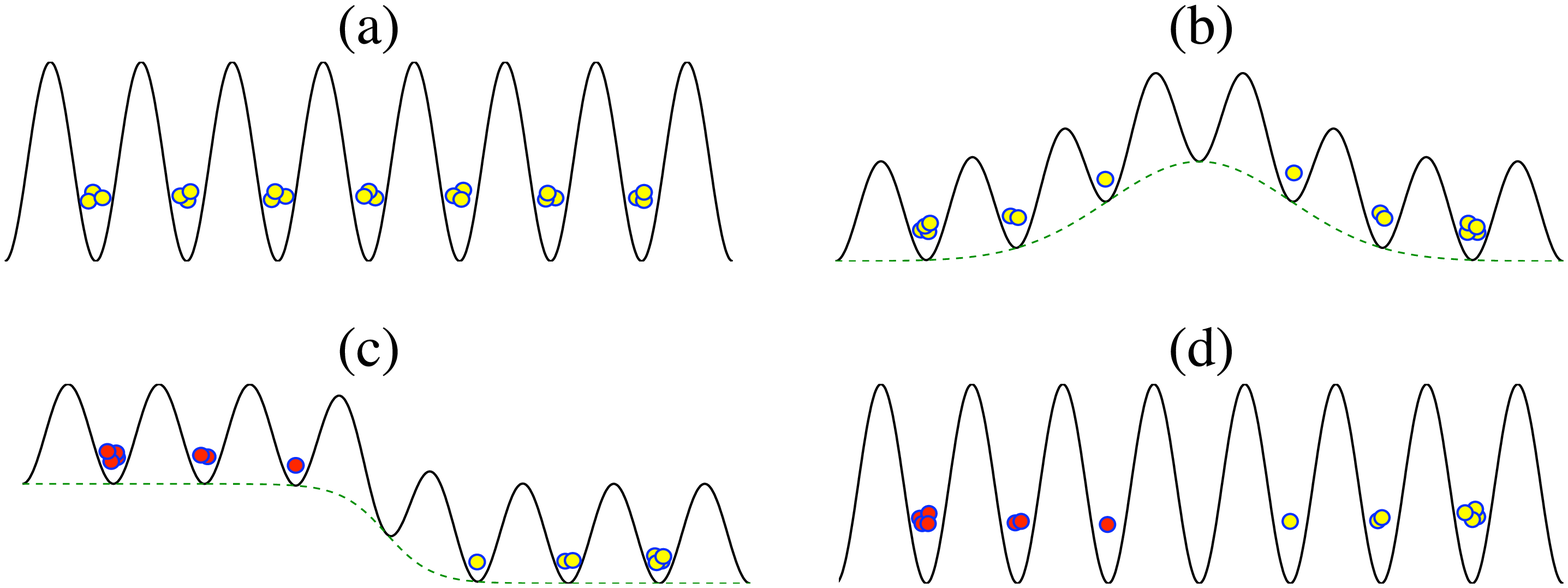}}}
\end{center}
\vspace{-0.25in}
\caption{(Color online) \emph{Density and phase engineering a standing soliton.}  To engineer a standing soliton in a lattice, we (a) take a free lattice, (b) apply a blue-detuned Gaussian beam to dig a density notch, (c) turn off the Gaussian beam and raise half the lattice to imprint a phase, and (d) lower the lattice to its original state.  The lattice heights are the same in each case, the number of atoms indicates the density, and red indicates atoms lower in phase by amount $\pi$. \label{fig:eng_standing_schematic}}
\end{figure}

In general, the unprecedented tunability permitted by optical lattice systems gives both theorists and experimentalists alike an ideal setting for investigating quantum many-body phenomena in a controlled fashion.  In these systems, experimentalists can adjust the height of the lattice by changing the intensity of the laser field used to create it.  With this knob, one can vary the ratio of the interaction to kinetic energy in the governing lattice Hamiltonian, e.g. the Bose-Hubbard Hamiltonian (BHH).  In fact, the first major experimental result for BECs in optical lattices~\cite{Greiner02_Nature_415_39} used just such a carefully controlled increase in optical lattice height past a critical value to demonstrate the Mott-superfluid quantum phase transition in the BHH~\cite{Fisher89_PRB_40_546, Jaksch98_PRL_81_3108}.  More generally, at present, experimentalists also have precise control over the lattice site filling factor, lattice geometry and dimensionality, species type, and atom-atom interaction strength, symmetry, and sign; the latter properties are due to use of a Feshbach resonance.  The system is also well-isolated from the environment, greatly curbing the intervention of finite-temperature effects.  All of these appealing properties make it possible to investigate out-of-equilibrium quantum many-body dynamics governed by the corresponding Hamiltonian, an aspect of the problem on which we focus in this paper and one that received little attention before the advent of optical lattices.  Specifically, we study how quantum many-body phenomena, e.g., quantum fluctuations and quantum entanglement, affect the behavior of dark solitons formed by ultracold bosons loaded into one-dimensional optical lattices.

Regardless of the physical system in which they occur, solitons are characterized by stable, non-dispersing propagation of either a minima (dark solitons) or a maxima (bright solitons) of the physically relevant dependent variable, e.g., particle number density in BECs.  It has been well-known for decades, since the seminal papers by Tsuzuki on nonlinear waves in the GP equation~\cite{Tsuzuki70_JLowTempPhys_4_441} and by Zakharov and Shabat on the inverse scattering transform~\cite{ZakharovShabat71, ZakharovShabat73}, that solitons exist as solutions to the one-dimensional (1D) nonlinear Schr\"{o}dinger equation (NLS).  It is also well-established that in the mean-field limit Bose-Einstein condensates (BECs) are well-described by the NLS~\cite{Dalfovo99_RevModPhys_71_463}.  The success of the GP equation as a model to describe the ultracold atomic Bose gas, as well as the observation of solitons in systems such as photonic crystals and nonlinear waveguide arrays, has excited a broad new interest in solitons in the last decade.  Both dark~\cite{Burger99_PRL_83_5198, Denschlag00_Science_287_97} and bright~\cite{Strecker02_Nature_417_150, Salomon02_Science_296_1290} solitons were observed rather early on in systems of BECs in harmonic trap geometries, and, due to ever improving experimental technique, there have very recently been exciting new developments regarding the creation and evolution of dark solitons formed by BECs~\cite{Sengstock08_NaturePhys_4_496, Sengstock08_PRL_101_120406, Oberthaler08_PRL_101_130401}.

To treat solitons in the context of a lattice via mean-field theory, there are two approaches:  (1) a full solution of the NLS with an external lattice potential~\cite{Bronski01_PRL_89_1402}, in which case the lattice soliton solutions of the NLS have been mapped out in detail~\cite{Louis03_PRA_67_013602, Efremidis03_PRA_67_063608}, or (2) a finite-element discretization of the NLS obtained by projecting the condensate wave function onto a basis of localized wave functions at each lattice site~\cite{Trombettoni01_PRL_86_2353} thus resulting in the discrete nonlinear Schr\"{o}dinger equation (DNLS).  Discrete lattice solitons have been studied using the DNLS both in purely mathematical~\cite{Kivshar94_PRE_50_5020, Johansson99_PRL_82_85} and BEC-related~\cite{Trombettoni01_PRL_86_2353, Kevrekidis03_PRA_68_035602} contexts.  On the experimental side, bright gap solitons have been observed in a system of repulsive \Rb~atoms trapped in a weak optical lattice~\cite{Eiermann04_PRL_92_230401}.  We emphasize that we will treat dark solitons with a size much greater than the lattice spacing, not gap solitons deep in the band gap which are localized on a few sites.

In contrast to these mean-field approaches, we use the BHH, a fully quantum model.  The DNLS emerges as the equation of motion when taking the mean-field limit of the BHH~\cite{Amico98_PRL_80_2189}.  This is why we are able to make quantitative predictions regarding the accuracy of using mean-field theory to describe soliton dynamics.  Related studies regarding the effect of quantum fluctuations on excited condensates in continuous geometries, e.g., dark solitons, using few-mode approximations and BDG-based methods can be found in the works of Dziarmaga, Sacha, and Karkuszewski~\cite{Dziarmaga02_PRA_66_043615,
Dziarmaga02_PRA_66_043620, Dziarmaga03_JPhysB_36_1217,
Dziarmaga06_JPhysB_39_57, Dziarmaga04_PRA_70_063616}, as we have already alluded to.  Using related methods, Law studied the dynamical depletion of dark solitons created via phase imprinting~\cite{Law03_PRA_68_015602}.  Below, we make connections with these works in our study.  An early study of quantum \emph{lattice} solitons is presented in~\cite{Scott94_PhysD_78_194}; a quantum theory of solitons in optical fibers was developed by Lai and Haus~\cite{Lai89_PRA_40_844}.

Thus, in conjunction with Ref. \cite{Mishmash09_short}, we present the first full entangled dynamical quantum many-body treatment of dark solitons formed by ultracold atoms in optical lattices.  In Sec.~\ref{sec:ModelsMethods}, we introduce the pertinent models used and briefly summarize the techniques used for solving them.  In Sec.~\ref{sec:Measures}, we define all of our measures for system characterization.  In Sec.~\ref{sec:QuantSoliEng}, we present results of quantum evolution of dark solitons obtained by density and phase engineering ground states of the BHH, a method in complete analogy to how dark solitons could be created in optical lattices experimentally.  In Sec.~\ref{sec:DNLSsolitons}, we consider quantum evolution in the BHH of mean-field solitons generated in the DNLS and add new measures and insight to the results in \cite{Mishmash09_short}.  In Sec.~\ref{sec:Bog}, we perform a Bogoliubov analysis of the dark soliton solutions in Sec.~\ref{sec:DNLSsolitons}. Finally, in Sec.~\ref{sec:ThatsAll}, we summarize our results.

\section{Models and Methods} \label{sec:ModelsMethods}

The governing model we employ throughout this paper is the single-band BHH~\cite{Fisher89_PRB_40_546, Jaksch98_PRL_81_3108} in one spatial dimension:
\begin{eqnarray}
\Hhat&=&-J\sum_{k=1}^{M-1}(\bhat_{k+1}^\dagger\bhat_k + \hc)\nonumber\\
&&+
\frac{U}{2}\sum_{k=1}^M\nhat_k(\nhat_k-\Ihat) +
\sum_{k=1}^M\epsilon_k\,\nhat_k, \label{eqn:BHH}
\end{eqnarray}
where $J$ is the nearest-neighbor hopping coefficient, $U$ is the on-site interaction coefficient, $\epsilon_k$ is an external potential aside from the lattice potential, and $M$ is the total number of lattice sites.  The destruction (creation) operator $\bhat_k$ ($\bhatd_k$) destroys (creates) a boson in the lowest Bloch-band Wannier state localized at lattice site $k$, while the number operator $\nhat_k\equiv\bhatd_k\bhat_k$ counts the boson occupation at site $k$.  We employ box (open) boundary conditions in all calculations, hence the restriction on the hopping term to contain only $M-1$ terms.

It is straightforward to derive Eq. (\ref{eqn:BHH}) from the continuous second-quantized many-body Hamiltonian describing a system of identical bosons with binary interactions after making the following assumptions: (1) the bosons interact with a short-range contact delta function potential; (2) the tight-binding approximation is valid so that next-nearest-neighbor hopping and nearest-neighbor interactions can be neglected; and (3) all relevant energy scales, e.g., temperature, hopping, and interaction energies, are smaller than the band spacing so that we can restrict the model to include only the lowest band.  This must be dynamically true as well as statically.

From this derivation, one can write down the coefficients $J$, $U$, and $\epsilon_i$ in terms of the single-particle Wannier functions on which Eq. (\ref{eqn:BHH}) is based~\cite{Jaksch98_PRL_81_3108}.  Equation (\ref{eqn:BHH}) is a truncation of the full continuous Hamiltonian that very accurately describes ultracold bosons trapped in optical lattices.  We note that the 1D model is obtainable experimentally from a full 3D lattice by increasing the strength of the lattice potential in the transverse directions, thus suppressing hopping in those directions.  This results in an array of 1D tubes with a lattice potential in the longitudinal direction.  This is not a true 1D quantum model though, in that we assume the transverse confinement length is still much larger than the range of the interatomic interaction potential and the actual atomic size.  That is, even though excitations can occur only along the longitudinal direction, the underlying scattering process is still three-dimensional, i.e., we are in the \emph{quasi-1D} regime.

As alluded to in Sec.~\ref{sec:Intro}, the DNLS can be obtained either as a finite-element discretization on the lattice of the continuous NLS~\cite{Trombettoni01_PRL_86_2353} or as a mean-field approximation of the BHH~\cite{Amico98_PRL_80_2189}.  In the latter case, it is assumed that the many-body state takes the form of a product of Glauber coherent states at each site $k$:
\begin{equation}
|\Psi\rangle = \bigotimes_{k=1}^M
|z_k\rangle,~~\mathrm{where}~~|z_k\rangle\equiv
e^{-|z_k|^2/2}\sum_{n=0}^\infty\frac{(z_k)^n}{\sqrt{n!}}|n\rangle.
\label{eqn:cohState1}
\end{equation}
Then, after taking the expectation value of Heisenberg's equation of motion, $i\hbar\partial_t\bhat_k=[\bhat_k,\Hhat]$, with $\Hhat$ given by Eq. (\ref{eqn:BHH}) but in normal ordered form, we arrive at the DNLS~\cite{Mishmash08_IMACS}:
\begin{equation}
i\hbar\dot{\psi}_k = -J\left (\psi_{k+1} + \psi_{k-1}\right) + U|\psi_k|^2
\psi_k + \epsilon_k \psi_k, \label{eqn:DNLS2}
\end{equation}
where $z_k=\langle\bhat_k\rangle\equiv\psi_k$ is the time-dependent coherent state amplitude at site $k$ that is equivalent to the condensate order parameter in the limit of a product of coherent states.  In Sec.~\ref{sec:DNLSsolitons}, we use a truncated form of the coherent states in Eq. (\ref{eqn:cohState1}) to build quantum analogs in the BHH of the dark soliton solutions of the DNLS, while in Sec.~\ref{sec:QuantSoliEng} we model our quantum soliton engineering in the BHH after an analogous calculation in the DNLS.  Finally, in Sec.~\ref{sec:Bog}, we treat the stability of DNLS soliton solutions using the Bogoliubov method.

For all simulations of the BHH in both real and imaginary time, we employ the time-evolving block decimation (TEBD) algorithm~\cite{Vidal03_PRL_91_147902, Vidal04_PRL_93_040502} that allows efficient classical simulation of the dynamics of slightly-entangled quantum states in 1D lattice Hamiltonians.  The main convergence parameter, denoted $\chi$, in this numerical method is the number of basis sets retained in a Schmidt decomposition at each bipartite splitting of the lattice.  For the specifics of our implementation of the algorithm, we refer the reader to Refs.~\cite{Mishmash08_IMACS, Mishmash08_MSThesis}.  The other approximations inherent in the algorithm are (1) a truncation of the local Hilbert space dimension at $d$ states so that each site can contain at most $d-1$ bosons and (2) a second-order Suzuki-Trotter representation of the time-evolution operator $e^{-i\Hhat\delta t/\hbar}$ over small time steps $\delta t$.  In all results presented herein, we have explicitly checked for convergence in $\chi$, $d$, and $\delta t$.  On the other hand, for all simulations of the DNLS in both real and imaginary time, we use a standard fourth-order Runge-Kutta routine~\cite{Mishmash08_MSThesis}.

\section{Measures} \label{sec:Measures}

We now clearly define the measures that we use to probe the physical properties of the system during time evolution.  The \emph{average particle number} $\langle\nhat_k\rangle$, i.e., the expectation value of the number operator at site $k$, is a local observable which corresponds directly to the particle density observed in experiment averaged over many runs.  Another local number-related observable is the \emph{normalized particle number variance} defined as $\eta_k\equiv\langle(\Delta n_k)^2\rangle/\langle\nhat_k\rangle\equiv
(\langle\nhat_k^2\rangle-\langle\nhat_k\rangle^2)/\langle\nhat_k\rangle$.  This measure characterizes the deviation of on-site number statistics away from the classical Poissonian limit.  Specifically, $\eta_k=1$ for atom-number Glauber coherent states which obey Poissonian number statistics, $\eta_k<1$ for a \emph{number-squeezed} state, $\eta_k>1$ for a \emph{phase-squeezed} state, and $\eta_k=0$ for a single Fock state of definite occupancy.  The \emph{order parameter} $\langle\bhat_k\rangle$, defined as the expectation value of the boson destruction operator at site $k$, corresponds directly to the DNLS solution $\psi_k$ in the limit of a direct product of atom-number Glauber coherent states.

Within our $d$-dimensional local Fock space, we define a Hermitian \emph{phase operator}, according to the Pegg-Barnett~\cite{Pegg89_PRA_39_1665} prescription as
\begin{eqnarray}
\thetahat_k&\equiv&\frac{(d-1)\pi}{d}\nonumber\\
&&+\frac{2\pi}{d}\sum_{n_k\neq
n_k'=0}^{d-1}\frac{|n_k'\rangle\langle n_k|}{\exp{[i(n_k'-n_k)2\pi/d]}-1},
\label{eqn:phaseop}
\end{eqnarray}
where we have set a reference phase $\theta_0$ to zero.  We use $\langle\thetahat_k\rangle$, the expectation value of the on-site phase operator, as a quantum measure of average relative phase between
lattice sites in characterizing quantum solitons.

The lattice representation of the \emph{single-particle density matrix} is an $M\times M$ matrix with elements defined as $(\rho_\mathrm{sp}^\mathrm{lat})_{ij} \equiv\langle\bhat^\dagger_j\bhat_i\rangle$, which can be used to calculate the system's \emph{single-particle natural orbitals} and \emph{quantum depletion}. The $(j+1)^\mathrm{th}$ most highly occupied natural orbital is computed as the $(j+1)^\mathrm{th}$ largest eigenvector of $\rho_\mathrm{sp}^\mathrm{lat}$.  The $k^\mathrm{th}$ component of this eigenvector, denoted $\phi_k^{(j)}$, is the coefficient of the natural orbital associated with site $k$; the eigenvalues of $\rho_\mathrm{sp}^\mathrm{lat}$ are the average occupation numbers of the natural orbitals and are denoted by $N_j$ with $N_0\geq N_1\geq\dots\geq N_{M-1}$.  We assume eigenvectors normalized to unity:  $\sum_{k=1}^M|\phi_k^{(j)}|^2=1$.  The quantum depletion is the proportion of particles \emph{not} in the condensed mode $\phi_k^{(0)}$:  $D\equiv 1-N_0/N$, where $N=\tr(\rho_\mathrm{sp}^\mathrm{lat})$ is the total number of particles in the system.

The \emph{pair correlation function}, defined as  $g^{(2)}_{ij}\equiv\langle\hat{b}_i^\dagger\hat{b}_j^\dagger\hat{b}_j\hat{b}_i\rangle$, is a measure of the joint probability that a particle will be measured at site $i$ and another particle will be measured at site $j$.  For visualization, we plot $g^{(2)}_k$, defined as the pair correlation between the center lattice site and site $k$ (for the case of an even number of sites we take the average of the pair correlation between site $k$ and the two centermost sites).

Out-of-equilibrium time evolution of bosons in an optical lattice is an ideal medium in which to study quantum entanglement dynamics in a many-body system.  We consider two types of entanglement in characterizing a system of atoms on a Bose-Hubbard lattice: entanglement of modes and entanglement of particles.  In the former case, we can think of each localized Wannier function as a quantum system that is, in general, entangled with the other $M-1$ Wannier functions.  Because different Wannier functions are spatially distinct, we can refer to entanglement between Wannier modes, i.e., sites, as \emph{spatial entanglement} or \emph{entanglement between sites}.  On the other hand, the particles themselves are interacting with one another so that the state of a given particle is, in general, entangled with the remaining $N-1$ particles~\cite{Einstein35_EPRpaper}.  We refer to this type of entanglement as \emph{particle entanglement}.  For each type of entanglement, we consider the average impurity $Q$ and the von Neumann entropy $S_\mathrm{vN}$.  For the case of spatial entanglement, we define the \emph{average local impurity}~\cite{Brennen03_QuantInfoAndComp_3_619} as
\begin{equation}
Q_\mathrm{modes}\equiv \frac{d}{d-1} \left [1 - \frac{1}{M}\sum_{k=1}^M \tr(\rhohat_k^2) \right
] \in [0, 1], \label{eqn:Qmeasure}
\end{equation}
where $\rhohat_k$ is the reduced density matrix describing site $k$.  The measure $Q_\mathrm{modes}$ quantifies multipartite entanglement in the sense that it averages the bipartite entanglement between each site and the remaining sites.  We note that the average local impurity is equally well described as the average local mixedness, a particular case of the generalized entropy~\cite{Barnum03_PRA_68032308, Barnum04_PRL_92_107902, Hu06_JMathPhys_47_023502}, or a local average quantum Tsallis entropy
\cite{Tsallis88_JStatPhys_52_479}.  On the other hand, the \emph{local von Neumann entropy} defined as
\begin{equation}
S_{\mathrm{vN},k} \equiv -\tr\left[\rhohat_k \log_d(\rhohat_k) \right ] \in [0,
1], \label{eqn:vNent}
\end{equation}
measures the entanglement between site $k$ and all other sites.  The analogous quantities for particle entanglement are the \emph{single-particle impurity},
\begin{equation}
Q_\mathrm{particles} \equiv \frac{M}{M-1}\left[1 -
\frac{1}{N^2}\tr(\rhohat_\mathrm{sp}^2)\right] \in [0, 1], \label{eqn:partImp}
\end{equation}
and the \emph{single-particle von Neumann entropy},
\begin{equation}
S_\mathrm{vN,\,particles} \equiv -\tr\left[\frac{\rhohat_\mathrm{sp}}{N}
\log_M\left(\frac{\rhohat_\mathrm{sp}}{N}\right) \right] \in [0, 1],
\label{eqn:partEntropy}
\end{equation}
where $\rhohat_\mathrm{sp}$ is the single-particle density matrix: $\langle i|\rhohat_\mathrm{sp}|j\rangle=(\rho_\mathrm{sp}^\mathrm{lat})_{ij}$ with $|i\rangle$ the lowest-band single-particle Wannier state centered at site $i$.

\section{Quantum Soliton Engineering} \label{sec:QuantSoliEng}

In this section, we present results obtained by using TEBD to simulate the generation and evolution of dark solitons in the BHH.  To create the dark soliton initial conditions, we generalize the methods of density and phase engineering originally applied to the continuous mean-field NLS~\cite{Carr01_PRA_63_051601} to the discrete mean-field DNLS and discrete quantum BHH.  Comparisons are made between soliton dynamics in the DNLS and BHH.  For all BHH simulations in both real and imaginary time, we employ a TEBD code optimized to conserve total particle number~\cite{Daley05_PhDThesis}.

Density engineering in a system described by either the DNLS or BHH can be accomplished as follows.  To mold the desired density, one selects an appropriate external potential $\epsilon_k$ and then finds the ground state of the resulting model, which we carry out in practice through imaginary time evolution. On the other hand, phase engineering is more subtle.  In mean-field theory, all atoms are assumed to be in the same wave function and hence in phase with one another.  Thus, imprinting a phase on the many-body wave function is conceptually equivalent to doing so on a single-particle wave function. Quantum-field theory, e.g., the BHH, relaxes the assumption that all particles are in-phase. However, we can still imprint a phase in the natural way so long as the condensate fraction is large enough.  To do this, we can change the local lattice potential to $\epsilon_k=-\theta_k\hbar/t_\mathrm{pulse}$ for a short time $t_\mathrm{pulse}$ in real time to imprint a phase $\theta_k$ at site $k$. In simulation, it suffices to apply the phase instantaneously.  This is accomplished by applying the on-site unitary operator $\exp(i\nhat\theta_k)$ to site $k$, where $\nhat$ is the local number operator.  In the DNLS, applying an instantaneous phase amounts to multiplying the DNLS wave function $\psi_k$ by the phase factor $e^{i\theta_k}$.  When applying the phase instantaneously, we are assuming that the density is unaltered during phase engineering; this requires that $t_\mathrm{pulse}$ be much less than the correlation time $\hbar/\mu$, where $\mu$ is the associated chemical potential. The exact forms of the external potential $\epsilon_k$ used for density engineering and of the imprinted phase $\theta_k$ depend on the problem in question.  The forms used for soliton engineering are given in the discussion that follows.

\begin{figure}
\begin{center}
\hspace{-0.1in}
\subfigure{\scalebox{0.21}{\includegraphics[angle=0]{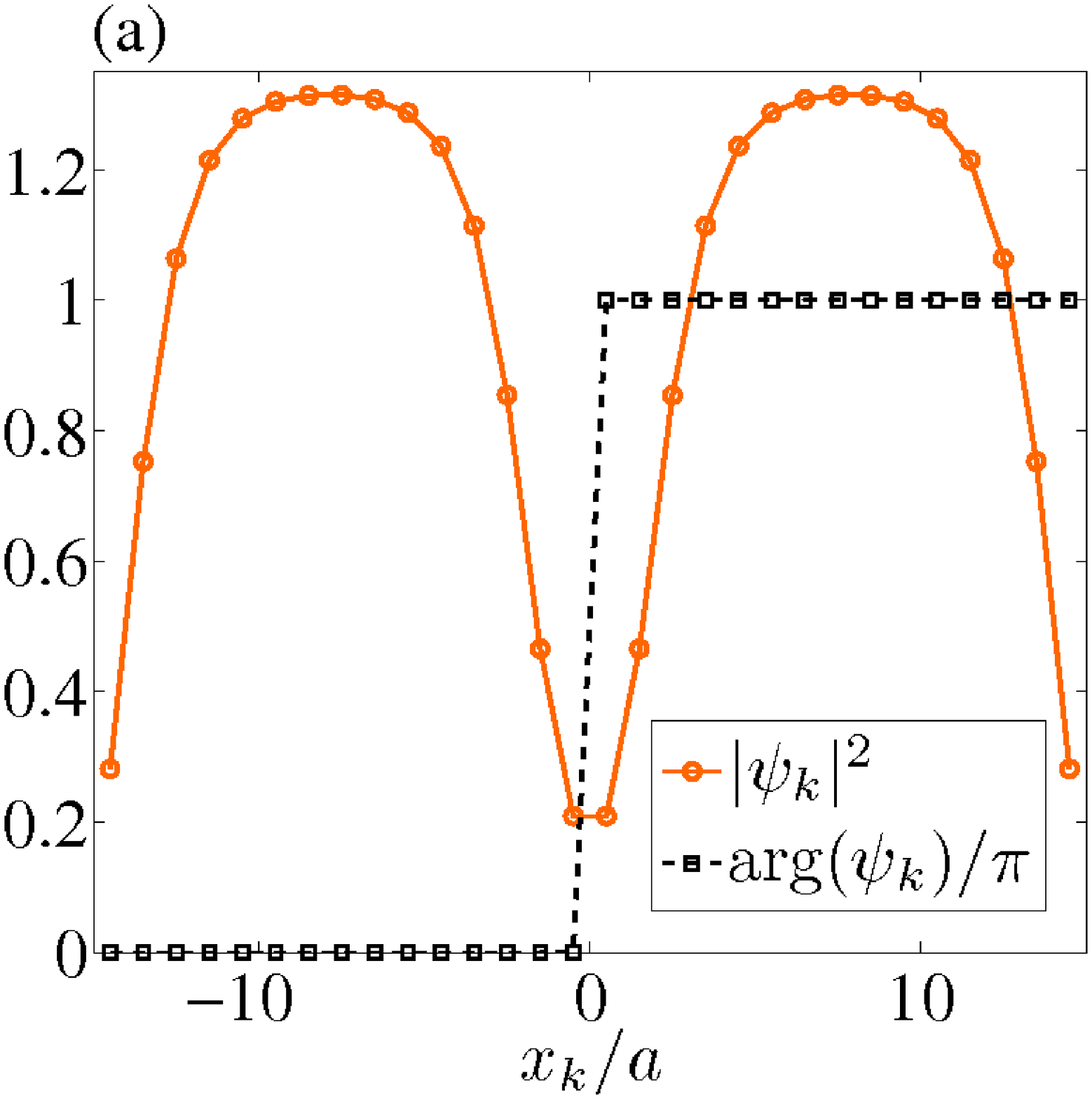}}}
\hspace{-0.0in}
\subfigure{\scalebox{\surfacePlotScale}{\includegraphics[angle=0]{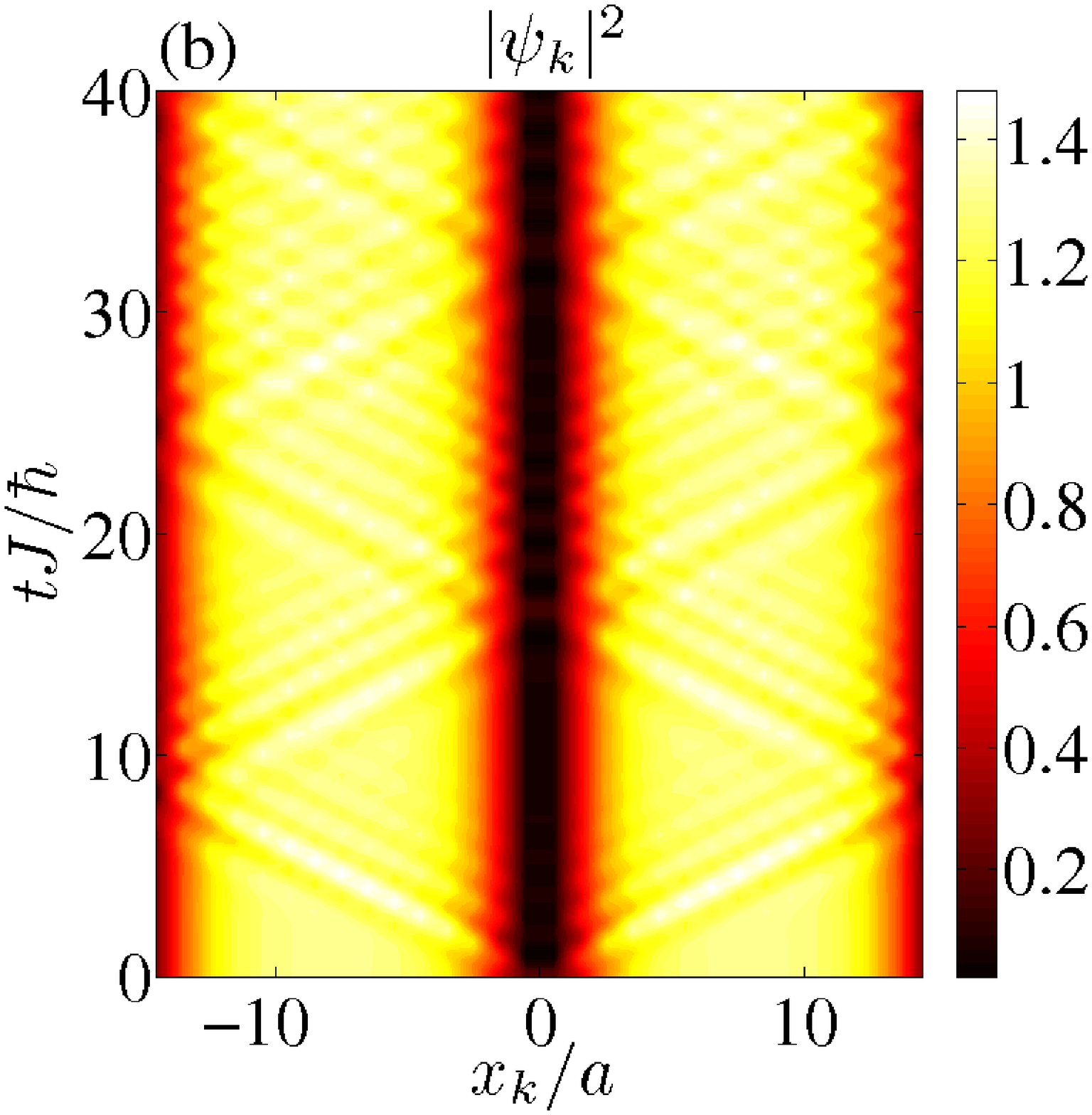}}}
\end{center}
\vspace{-0.25in}
\caption{(Color online) \emph{Engineered standing soliton profile and DNLS dynamics.} (a) The density and phase of the engineered standing DNLS soliton solution is plotted versus spatial position.  Because we select an even number of sites, the two points with lowest density straddle the lattice center at $x=0$.  (b)  Real time dynamics in the DNLS of the initial condition in (a) show a stably propagating density notch with a spray of phonons.  A plot of the phase (not shown) reveals a $\pi$ phase drop across the notch for all times.  Note that all surface plots in this paper have space on the horizonal axis and time on the vertical axis. \label{fig:eng_standing_DNLS}}
\end{figure}

We first concentrate on the case of an engineered standing soliton in the center of the lattice (see Fig.~\ref{fig:eng_standing_schematic}).  A single Gaussian beam is used to engineer the density notch and a hyperbolic tangent phase profile is then imprinted across the notch.  We can dig a single density notch at the lattice center, i.e., at $x=0$, by performing imaginary time relaxation with an external Gaussian potential of the form
\begin{equation}
\epsilon_k = V_1\exp\left(-\frac{x_k^2}{2\sigma_{V1}^2}\right).
\label{eqn:gaussPot1}
\end{equation}
For the phase imprinting, we use a hyperbolic tangent phase profile of the form
\begin{equation}
\theta_k = \frac{\Delta\theta_1}{2}\left[\tanh\left(\frac{2x_k}{\sigma_{\theta
1}}\right)+1\right], \label{eqn:phaseImprint1}
\end{equation}
where in each case $x_k/a=k-(M+1)/2$ is the position of site $k$ with $a$ the lattice constant.

For a true standing dark soliton, we would have to use a Dirac delta function for the potential in Eq. (\ref{eqn:gaussPot1}) to effectively create two independent boxes so the wave function would heal perfectly to zero at the center of the lattice.  Thus, for a Gaussian potential of finite amplitude and width, a certain fraction of the energy used to form the notch goes into the creation of phonons and a certain fraction goes into the creation of the soliton~\cite{Carr01_PRA_63_051601}.  Also, standing solitons have a $\pi$ phase drop across the notch in the form of a perfect step function; hence, ideally we should take $\Delta\theta_1=\pi$ and $\sigma_{\theta 1}\rightarrow 0$. Experimentally, the parameter $\sigma_{\theta 1}$, i.e., the width of the phase profile, represents diffraction-limited fall-off of the far-detuned laser beam used to raise half the lattice~\cite{Carr01_PRA_63_051601}.  If $\sigma_{\theta 1}$ is on the order of the soliton healing length, the engineered soliton will drift to the left. We ignore this effect in simulation by taking a sufficiently small value of $\sigma_{\theta 1}$.  Note, however, that stationary solitons can still be created with significant diffraction-limited fall-off by choosing a value of $\Delta\theta$ slightly greater than $\pi$ to create a second shallow soliton to carry away the
momentum of the first~\cite{Carr01_PRA_63_051601}.

Density and phase engineering using the forms given by Eqs. (\ref{eqn:gaussPot1}) and (\ref{eqn:phaseImprint1}) can be applied to either the DNLS or the BHH for soliton creation.  We consider both cases in this section.  A schematic of the procedure used to quantum engineer a stationary dark soliton is depicted in Fig.~\ref{fig:eng_standing_schematic}.

Let us now consider a single simulation of soliton engineering in both the mean-field (DNLS) and quantum many-body (BHH) pictures.  We choose to work on a lattice of $M=30$ sites with $N=30$ particles so that the filling factor $\nu\equiv N/M=1$.  The effective interaction parameter is taken to be $\nu U/J=0.35$.  This is an important parameter because, regardless of the values of $\nu$ and $U/J$ separately, the DNLS statics and dynamics are equivalent, up to a normalization of the order parameter $\psi_k$, for fixed $\nu U/J$ at a given $M$.  The relevant parameters used to engineering the soliton are $V_1/J=1$, $\sigma_{V1}/a=1$, $\Delta\theta=\pi$, $\sigma_{\theta 1}/a=0.1$.  The reason that we select an even number of sites is that it allows us to set a reasonable amplitude $V_1$ for the Gaussian potential.  For an odd number of sites, $V_1$ would have to be increased substantially in order for a node to be created at $x=0$, a requirement for the dark soliton to be stationary.  Not increasing $V_1$ for an odd number of sites results in an instability of the soliton since the wave function does not vanish at the origin, an effect we have observed. The solitons created with this method with an even number of sites are similar to the ``$B$ modes" analyzed in Ref.~\cite{Johansson99_PRL_82_85}, but without the site-to-site staggered sign structure.

Figure~\ref{fig:eng_standing_DNLS}(a) depicts the soliton profile initial condition obtained in the DNLS after imaginary time evolution with the external potential (\ref{eqn:gaussPot1}) and application of the phase imprint (\ref{eqn:phaseImprint1}); the subsequent real time dynamics are shown in Fig.~\ref{fig:eng_standing_DNLS}(b).  We only show the density in Fig.~\ref{fig:eng_standing_DNLS}(b), while, as expected, the phase grows approximately linearly in time with a $\pi$ phase drop across the notch for all times.  Due to our use of a finite-size Gaussian for density engineering, phonons are also created.  We now wish to compare these results to a full quantum many-body calculation using TEBD.

\begin{figure}
\begin{center}
\subfigure{\scalebox{\surfacePlotScale}{\includegraphics[angle=0]{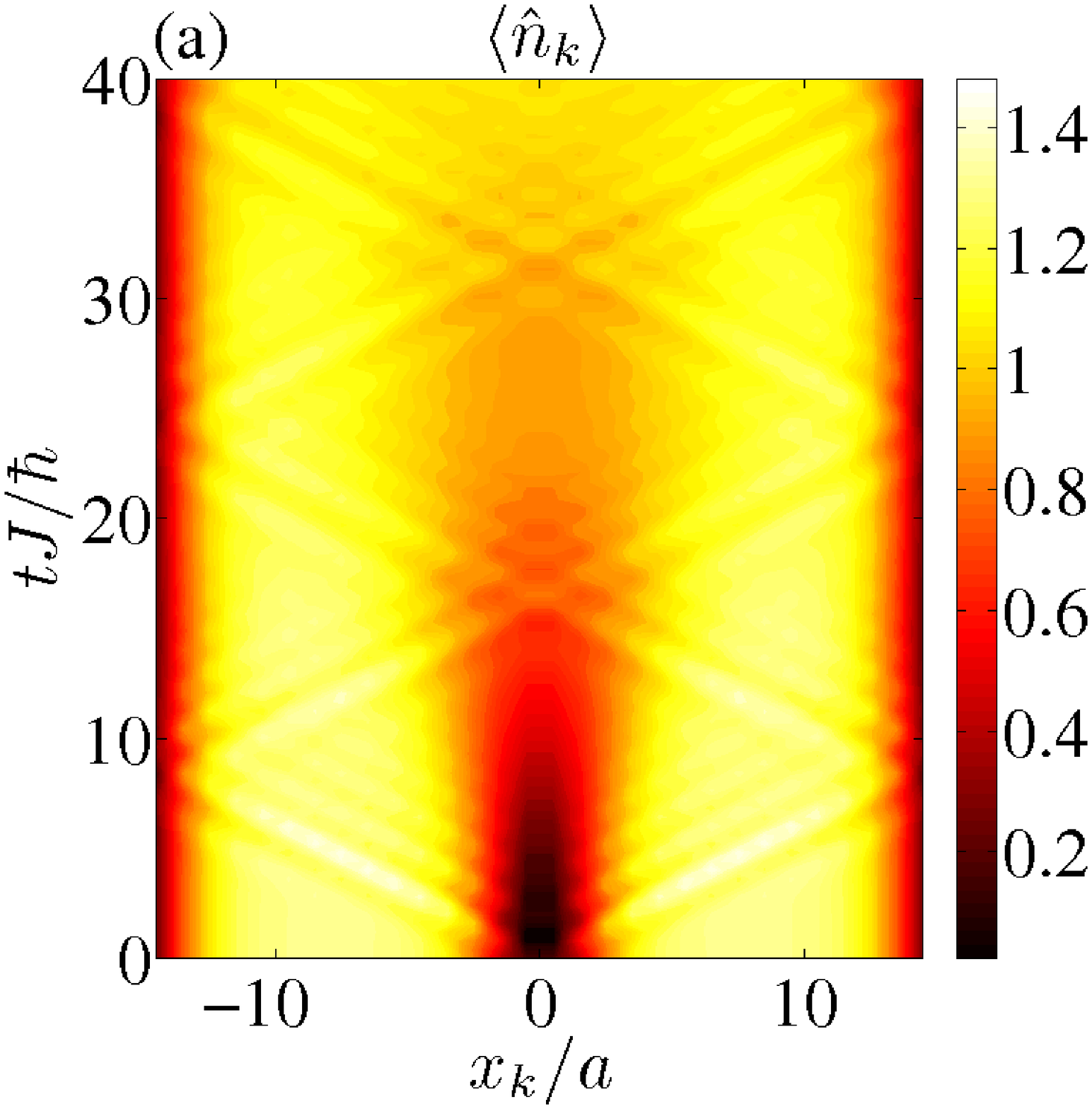}}}
\hspace{-0.0in}
\subfigure{\scalebox{\surfacePlotScale}{\includegraphics[angle=0]{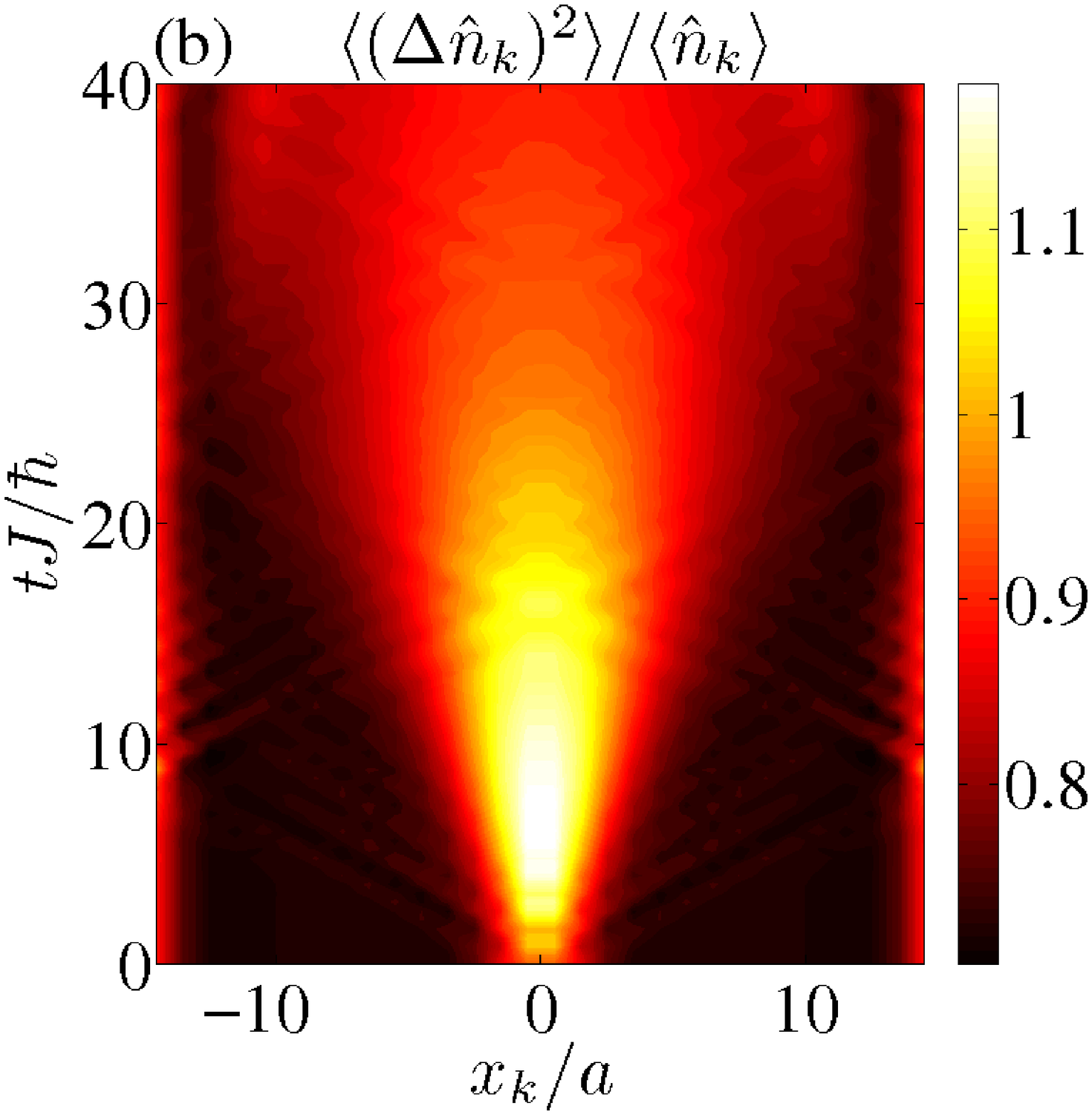}}}
\end{center}
\vspace{-0.25in}
\caption{(Color online) \emph{Average number and number fluctuations for engineered standing quantum soliton.}  (a)  The average particle density during quantum evolution indicates that the dark soliton fills in over time; $\langle\nhat_k\rangle$ can be thought of as the sum of all natural orbitals weighted by their respective occupation numbers, cf. Fig.~\ref{fig:eng_standing_natorbs}.  (b) The normalized number variance measuring the relative on-site number fluctuations is greatest as the soliton fills in. \label{fig:eng_standing_number}}
\end{figure}

\begin{figure}[t]
\begin{center}
\subfigure{\scalebox{\surfacePlotScale}{\includegraphics[angle=0]{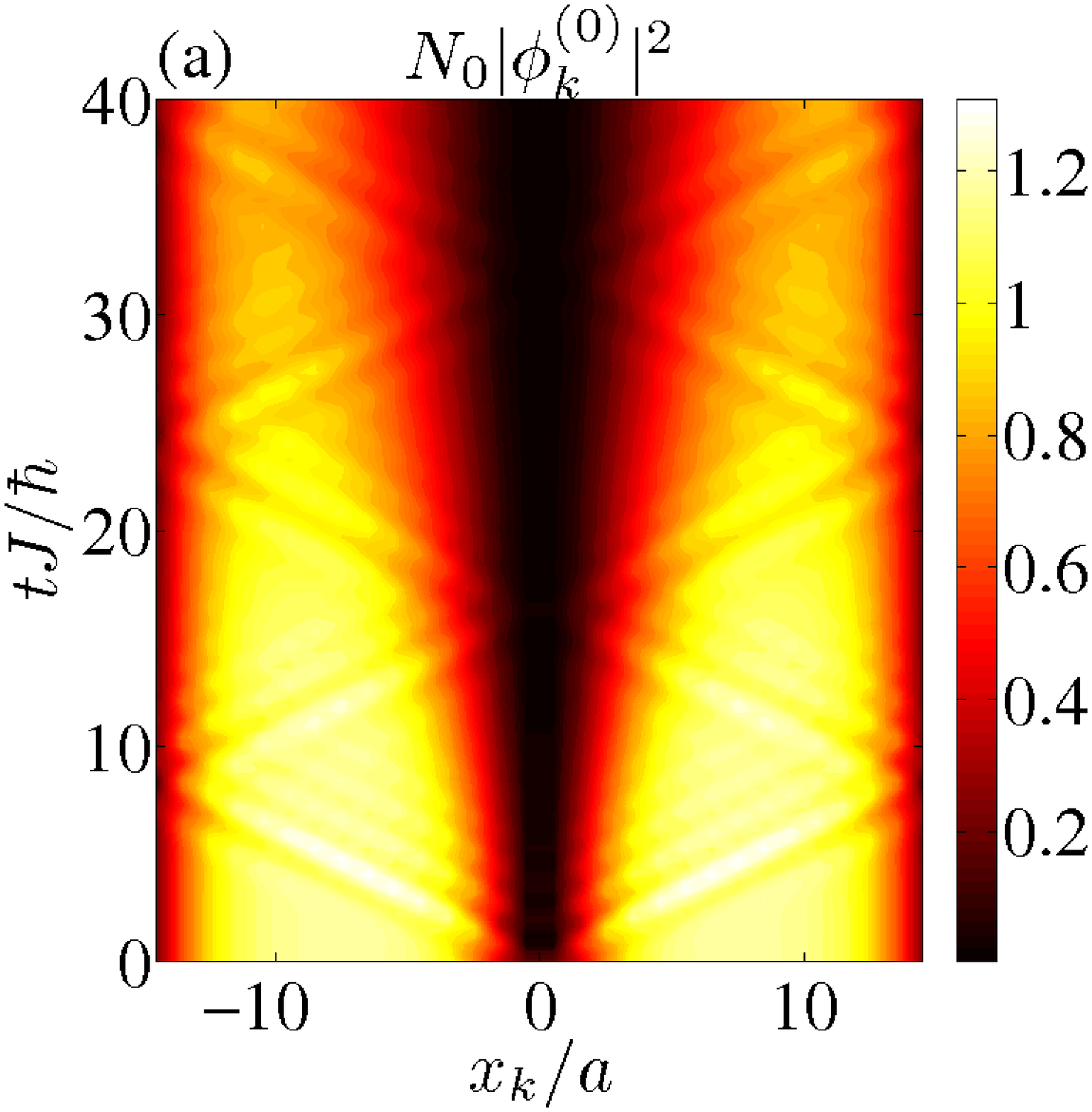}}}
\hspace{-0.0in}
\subfigure{\scalebox{\surfacePlotScale}{\includegraphics[angle=0]{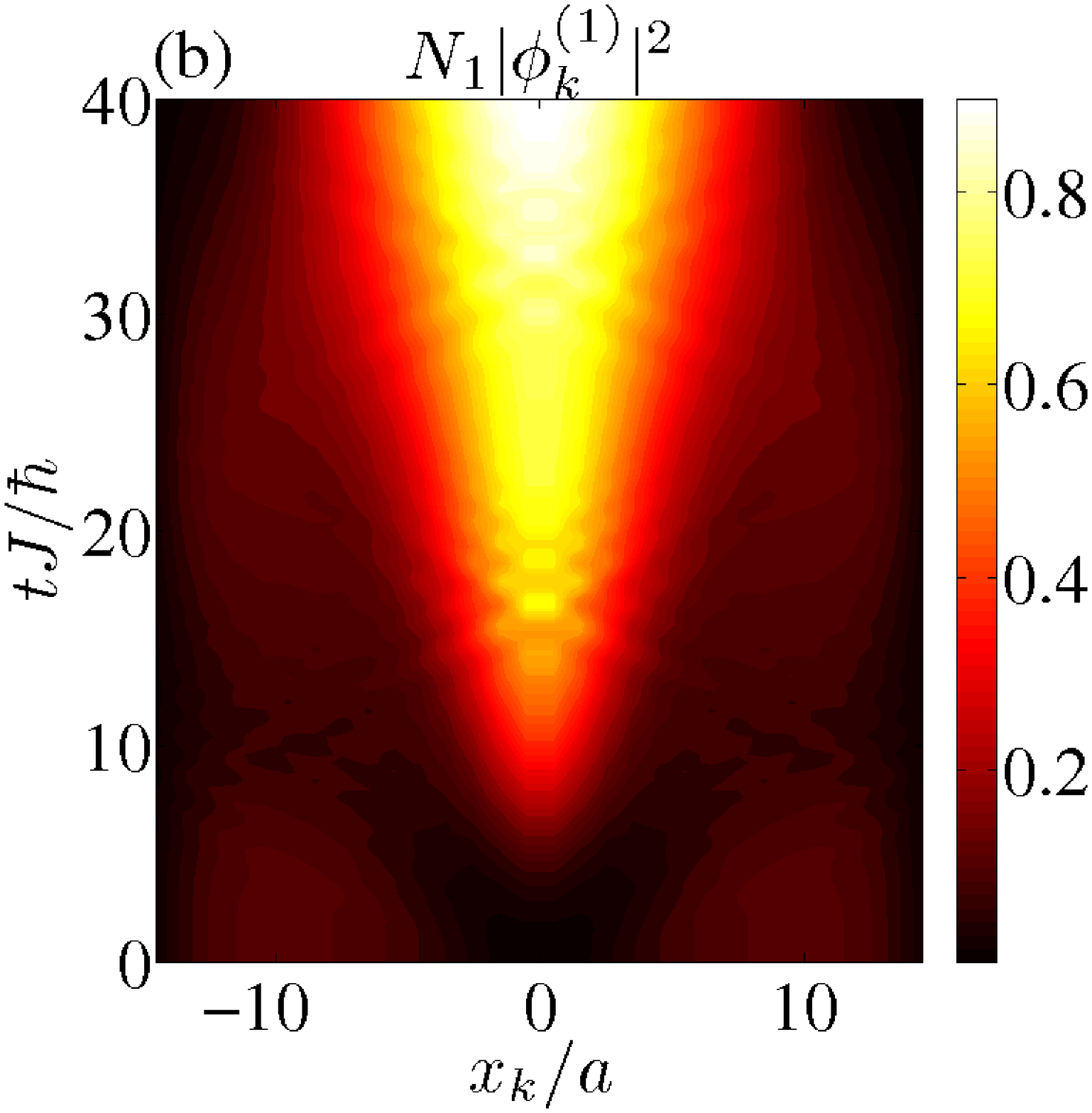}}}
\end{center}
\vspace{-0.30in}
\begin{center}
\subfigure{\scalebox{\surfacePlotScale}{\includegraphics[angle=0]{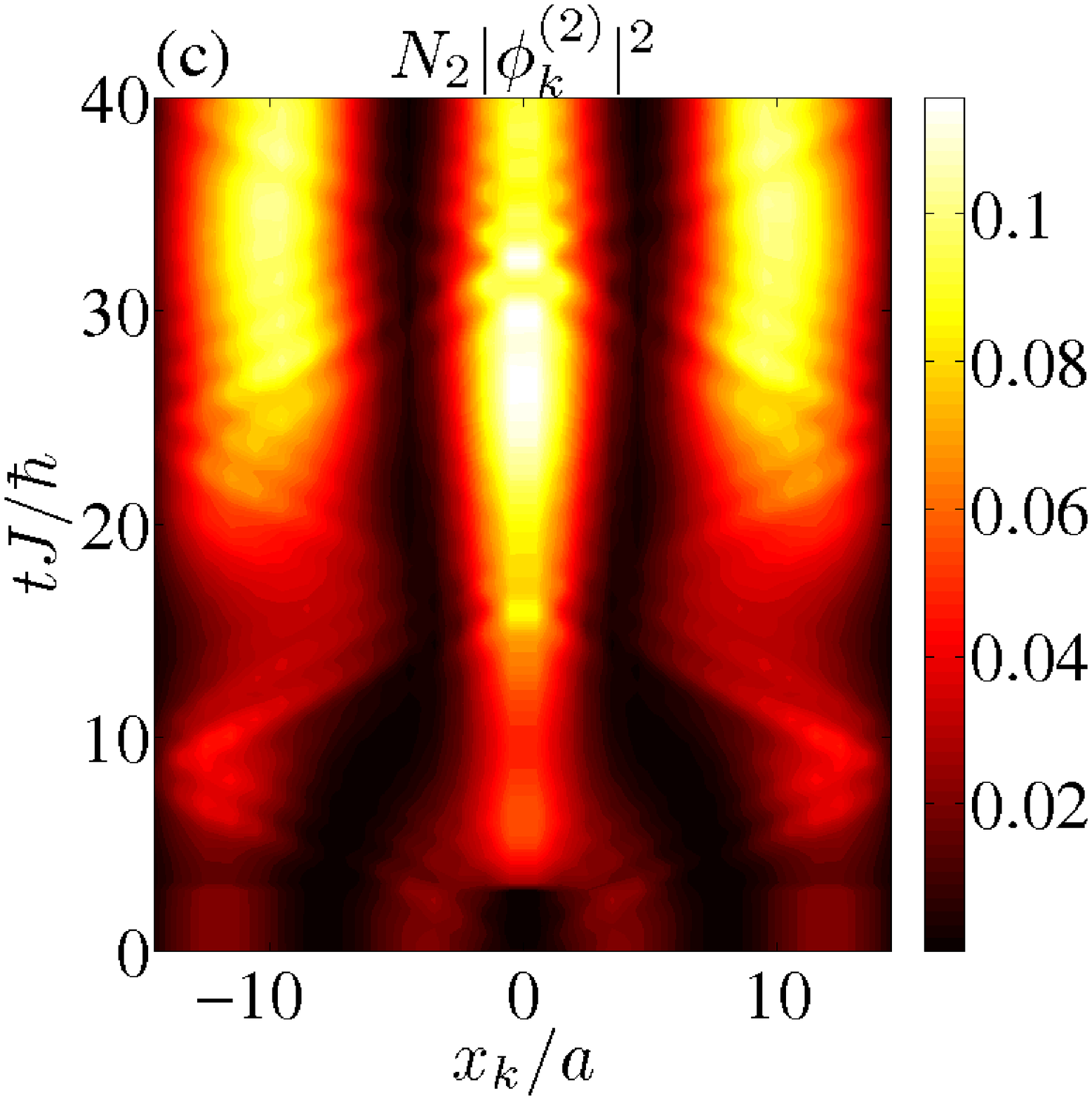}}}
\hspace{-0.0in}
\subfigure{\scalebox{\surfacePlotScale}{\includegraphics[angle=0]{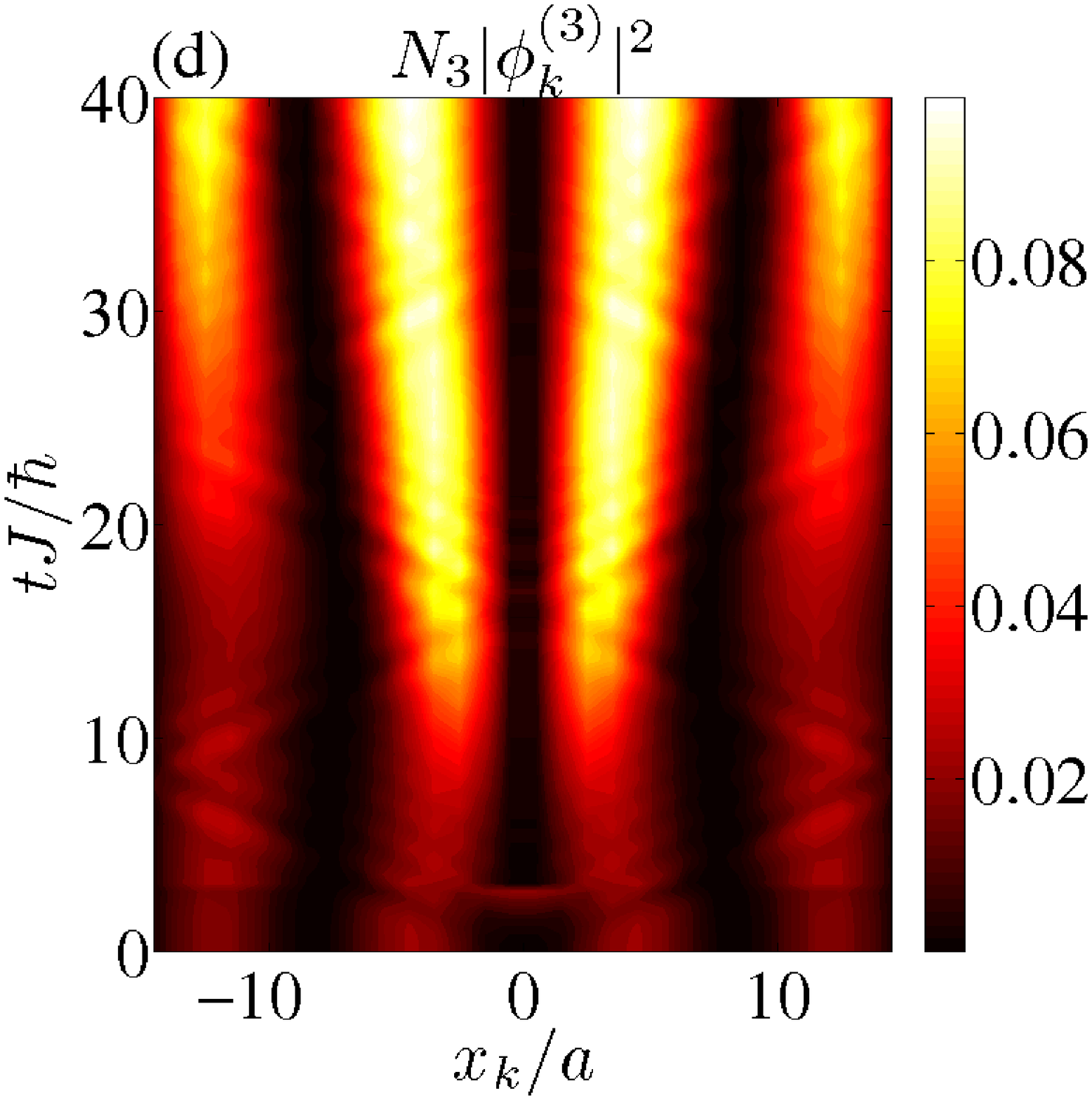}}}
\end{center}
\vspace{-0.25in}
\caption{(Color online) \emph{Natural orbital dynamics for engineered standing quantum soliton.}  The densities of the four most highly occupied natural orbitals are plotted in (a)--(d) in order of decreasing occupation $N_j$.  Population of the mode $\phi^{(1)}_k$ is most responsible for the soliton decay shown in Fig.~\ref{fig:eng_standing_number}(a).  Here, we show the natural orbital densities $|\phi^{(j)}_k|^2$ multiplied by their respective average occupation numbers $N_j$. \label{fig:eng_standing_natorbs}}
\end{figure}

\begin{figure}
\begin{center}
\subfigure{\scalebox{\surfacePlotScale}{\includegraphics[angle=0]{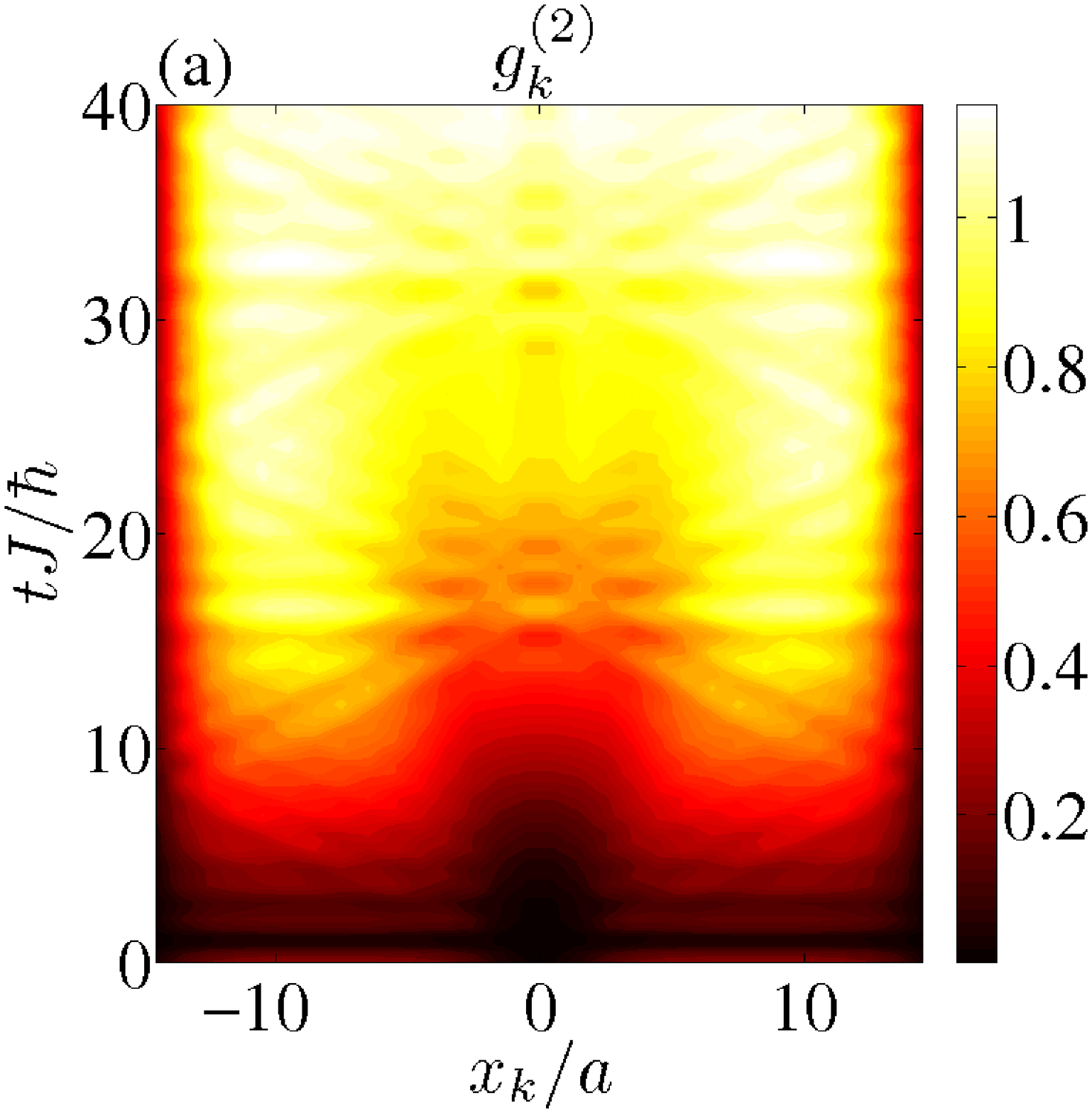}}}
\hspace{-0.0in}
\subfigure{\scalebox{\surfacePlotScale}{\includegraphics[angle=0]{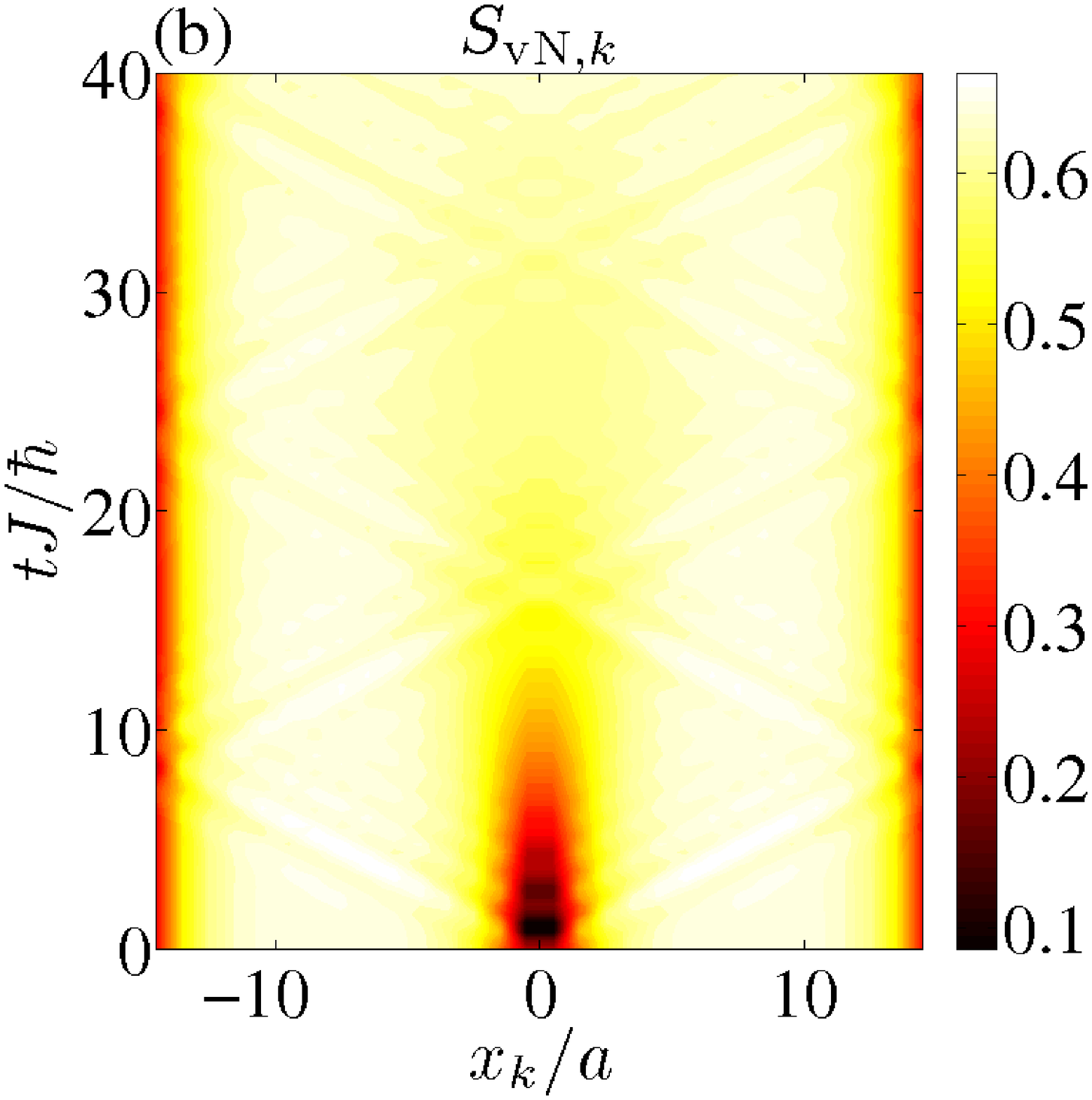}}}
\end{center}
\vspace{-0.30in}
\begin{center}
\subfigure{\scalebox{\surfacePlotScale}{\includegraphics[angle=0]{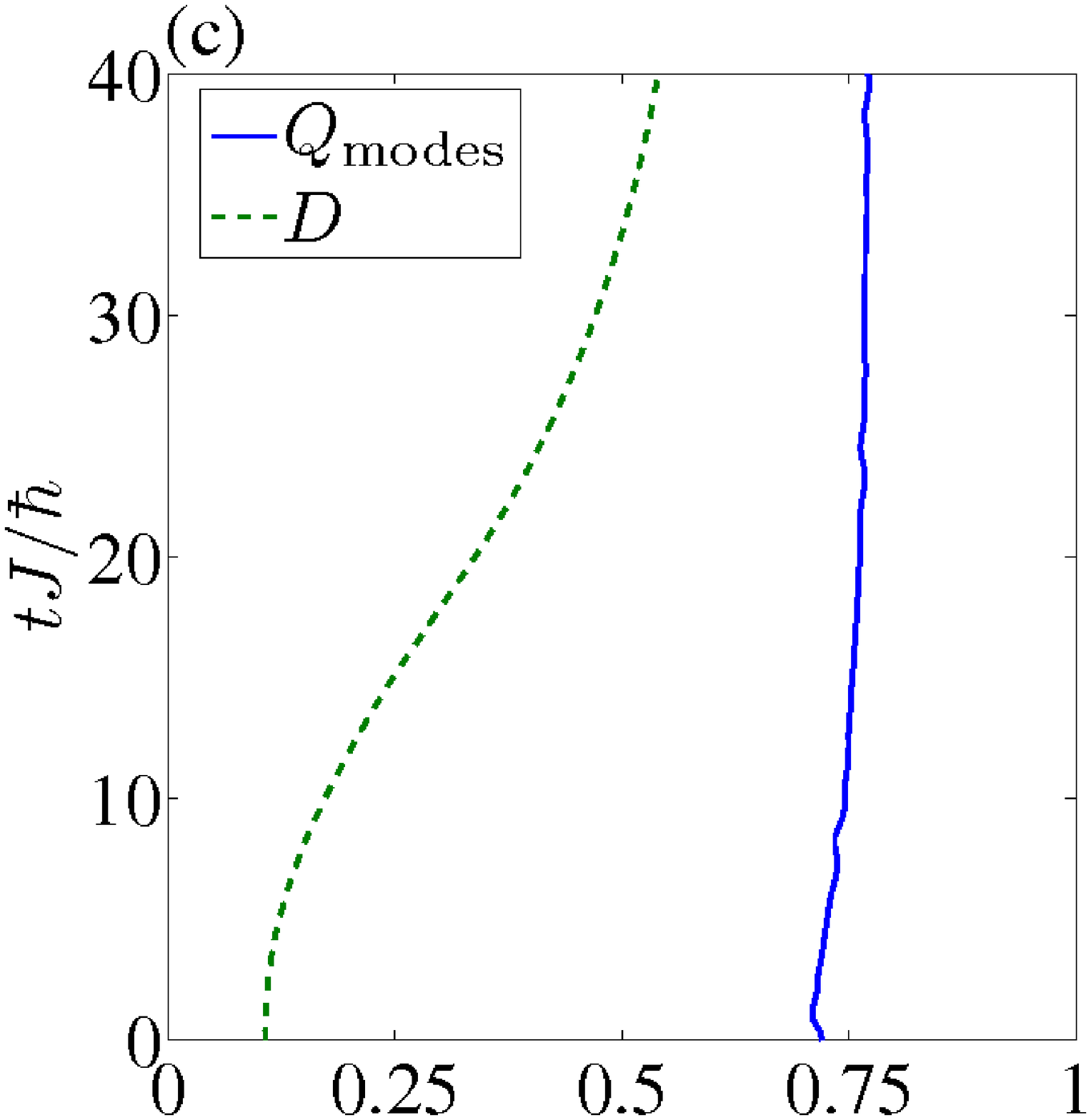}}}
\hspace{-0.0in}
\subfigure{\scalebox{\surfacePlotScale}{\includegraphics[angle=0]{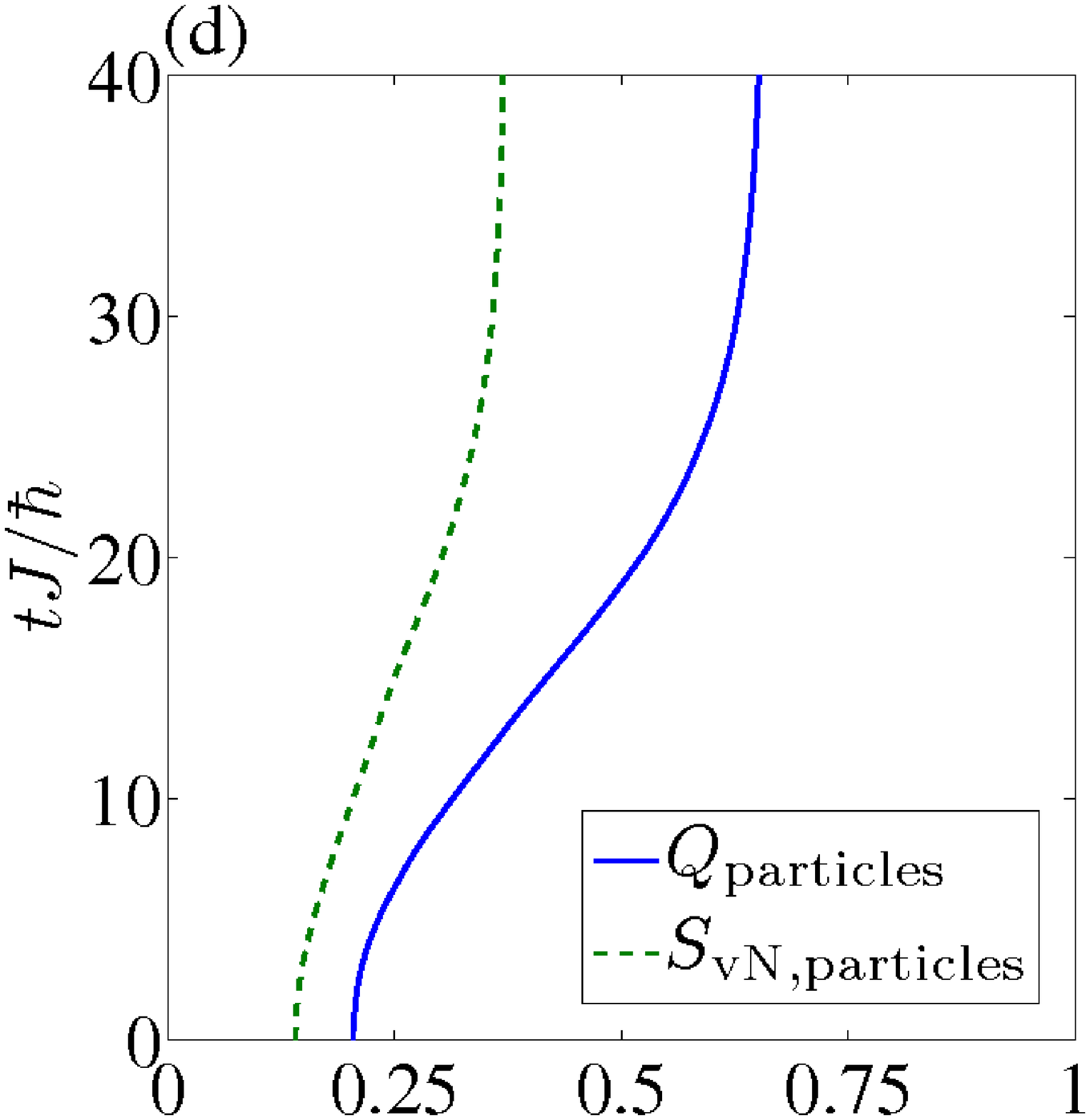}}}
\end{center}
\vspace{-0.25in}
\caption{(Color online) \emph{Quantum measures for engineered standing quantum soliton.} (a)  The pair correlation function with respect to the center lattice site tells us that the soliton is in fact filling in during a single experiment.  The initial condition has substantial spatial entanglement as indicated by (b) the local von Neumann entropy and (c) the average local impurity.  At short times, before the soliton has filled in, the entropy in the region of the soliton is lower than at other points in the lattice, but this notch in entropy fills in over time.  In (c) and (d), we see that both quantum depletion and particle entanglement grow in time.  \label{fig:eng_standing_quantum}}
\end{figure}

To this end, we take the same parameters as used for the DNLS and perform imaginary time propagation in TEBD with the same Gaussian potential used to produce the density structure in Fig.~\ref{fig:eng_standing_DNLS}(a). Retaining $\chi=120$ basis sets and allowing up to $d-1=7$ particles per site, we are able to calculate a ground state well-converged density-engineered ground state via TEBD imaginary time evolution. The quantum depletion in this ground state is only $D\approx 11\%$. We then remove the Gaussian potential and model the application of an instantaneous phase imprint by applying to each site $k$ the on-site unitary $\exp(i\nhat\theta_k)$, where $\theta_k$ is the same narrow tanh phase profile shown in Fig.~\ref{fig:eng_standing_DNLS}(a). The resulting quantum dynamics are shown in Figs.~\ref{fig:eng_standing_number},~\ref{fig:eng_standing_natorbs}, and~\ref{fig:eng_standing_quantum}.

The evolution of the average particle number density and normalized number variance can be seen in Fig.~\ref{fig:eng_standing_number}.  We see that, in contrast to the DNLS dynamics of Fig.~\ref{fig:eng_standing_DNLS}(b), time evolution according to the BHH causes the soliton to fill in.  This is qualitatively similar to the phenomena we have observed using mean-field DNLS soliton solutions as initial conditions for quantum evolution~\cite{Mishmash09_short, Mishmash08_IMACS}.  The normalized variance plot indicates that the relative number fluctuations are largest in the regions of space-time where the soliton is filling in.

The filling-in phenomena can be attributed to allowed two-body scattering processes that deplete particles out of the original solitonic condensate wave function into modes with nonvanishing density at the lattice center.  To see this explicitly, we plot the densities of the four most highly occupied natural orbitals in Fig.~\ref{fig:eng_standing_natorbs}.  We see in Fig.~\ref{fig:eng_standing_natorbs}(a) that the condensate wave function $\phi^{(0)}_k$ retains its solitonic form throughout time evolution but that occupation of higher order modes causes the soliton to fill in over time.  In the mean-field DNLS theory, the symmetry of the initial condition is preserved throughout time evolution in a manner very similar to that of the single-particle \schrod equation.  Hence, for an initial condition in the form of a dark soliton, the antisymmetry of the condensate wave function is preserved during mean-field evolution:  recall the persistent $\pi$ phase drop across the notch in the Fig.~\ref{fig:eng_standing_DNLS}(b).  Because mean-field theory assumes all bosons to be in the same configuration for all time, the soliton is a stable configuration of the DNLS.

However, as we have shown, a full quantum treatment such as simulation of the BHH predicts depletion out of the solitonic condensate wave function into modes that effectively fill in the notch.  Because derivation of the BHH takes into account only two-body scattering processes, it must be these processes that are responsible for depletion into non-solitonic orbitals.  The symmetry of the many-body wave function is preserved regardless of the model chosen.  Therefore, the mechanism most responsible for soliton graying should involve two atoms depleting from the antisymmetric condensate wave function into symmetric modes, the lowest energy of which is shown in Fig.~\ref{fig:eng_standing_natorbs}(b) and has significant density in the region of the soliton notch. The other main scattering process involves two atoms previously depleted into a symmetric mode scattering into an antisymmetric mode of higher order than the soliton mode. Both of these processes retain the symmetry of the many-body wave function.  Although the population of $\phi^{(1)}_k$ is most responsible for soliton dissipation, higher order modes also contribute, e.g., $\phi^{(2)}_k$ in Fig.~\ref{fig:eng_standing_natorbs}(c).

The time dependence of the \emph{phase} of the natural orbitals is difficult to show graphically because at each time step an independent diagonalization of the single-particle density matrix is performed, and the diagonalization routine arbitrarily assigns a global phase to the eigenvectors.  However, the symmetry of the natural orbitals can still be discerned from these plots regardless of their unsightliness.  For completeness, we quote the symmetries of the lowest four modes as follows:  $\phi^{(0)}_k$ is an antisymmetric mode, $\phi^{(1)}_k$ and $\phi^{(2)}_k$ are both symmetric, and $\phi^{(3)}_k$ is antisymmetric.

Similar conclusions are reached in Ref.~\cite{Dziarmaga02_PRA_66_043615} where the authors use a three-mode approximation to model the antisymmetric to symmetric depletion of a soliton in a continuous trap geometry.  Our treatment has the advantage that bosons are allowed to occupy any of the $M$ allowable natural modes of the lattice so that both depletion processes can occur. Our method also allows us to directly probe the quantum many-body nature of the system using, for example, the pair correlation function and entanglement measures as depicted in Fig.~\ref{fig:eng_standing_quantum}.

By calculating the average local density as in Fig.~\ref{fig:eng_standing_number}(a), we have only shown that the dark soliton fills in on average over many density measurements.  Thus, we cannot yet say what would happen during a single experiment:  does the soliton in fact fill in or does it wander randomly to the left and right without filling in at all?  We can address this question by examining the pair correlation function $g^{(2)}_k$ defined in Sec.~\ref{sec:Measures}.  If the soliton were wandering, this measure would vanish at values $x_k\neq 0$ even during the the later times of evolution when $\langle\nhat_k\rangle$ indicates that the soliton has filled in.  This is because $g^{(2)}_k$ is a measure of the joint probability for measuring particles at the origin (the initial position of the soliton) \emph{and} at site $k$.  But, for a wandering dark soliton, if a particle is measured at the origin, there would be another site, $k_0$ say, with vanishingly small density due to the presence of the displaced soliton.  This would imply $g^{(2)}_{k_0}$ to be approximately zero.  In Fig.~\ref{fig:eng_standing_quantum}(a), we depict the time dependence of the pair correlation function and show that, in fact, there is not some $k_0$ for which $g^{(2)}_{k_0}$ vanishes even during later times.  This result is one strong piece of evidence that our study is a full quantum entangled dynamical study of dark solitons; we present yet another piece of evidence in Fig.~\ref{fig:eng_colliding_number} when we discuss a collision between two dark solitons.

In Fig.~\ref{fig:eng_standing_quantum}, we also depict the dynamics of the entanglement measures defined at the end of Sec.~\ref{sec:Measures}.  The local entropy plot in Fig.~\ref{fig:eng_standing_quantum}(b) shows that the soliton engineering causes the local von Neumann entropy to decrease in the region of the density notch.  Time evolution causes this notch in entropy to dissipate.  In contrast to our results pertaining to quantum evolution of mean-field solitons discussed in Sec.~\ref{sec:DNLSsolitons} and Ref.~\cite{Mishmash09_short}, both the local von Neumann entropy and average local impurity have finite value at $t=0$; $Q_\mathrm{modes}$ remains approximately constant in time although it grows slightly while the soliton is filling in.  Panels (c) and (d) of Fig.~\ref{fig:eng_standing_quantum} depict this time dependence of $Q_\mathrm{modes}$ as well as growth in quantum depletion $D$,  single-particle impurity $Q_\mathrm{particles}$, and single-particle von Neumann entropy $S_\mathrm{vN,particles}$.

The behavior presented above for the quantum evolution of a density- and phase-engineered standing dark soliton is general.  The dependence of the dynamics on parameter choices can be summarized as follows.  Increasing the effective interaction strength $\nu U/J$ at fixed filling $\nu$ and lattice size $M$ causes the soliton to decay at a faster rate, i.e., the soliton lifetime decreases.  Also, increasing the filling factor at a fixed lattice size, but keeping the effective mean-field interaction $\nu U/J$ fixed, causes an increase in the soliton lifetime.
Since the effective strength of quantum fluctuations is quantified by $U/(2\nu J)$ in the 1D BHH~\cite{Kollath05_PRA_71_053606, Danshita09_PRA_79_043601}, these trends clearly indicate that the soliton decay shown above is due to quantum fluctuations.  A plot of these trends can be found in Fig. 2 of Ref. \cite{Mishmash09_short} where we use the above method of dark soliton engineering in the BHH to compute the soliton lifetime versus $\nu U/J$ at fixed filling factors $\nu=0.5,1.0,1.5,2.0$.

\begin{figure}
\begin{center}
\hspace{-0.1in}
\subfigure{\scalebox{0.21}{\includegraphics[angle=0]{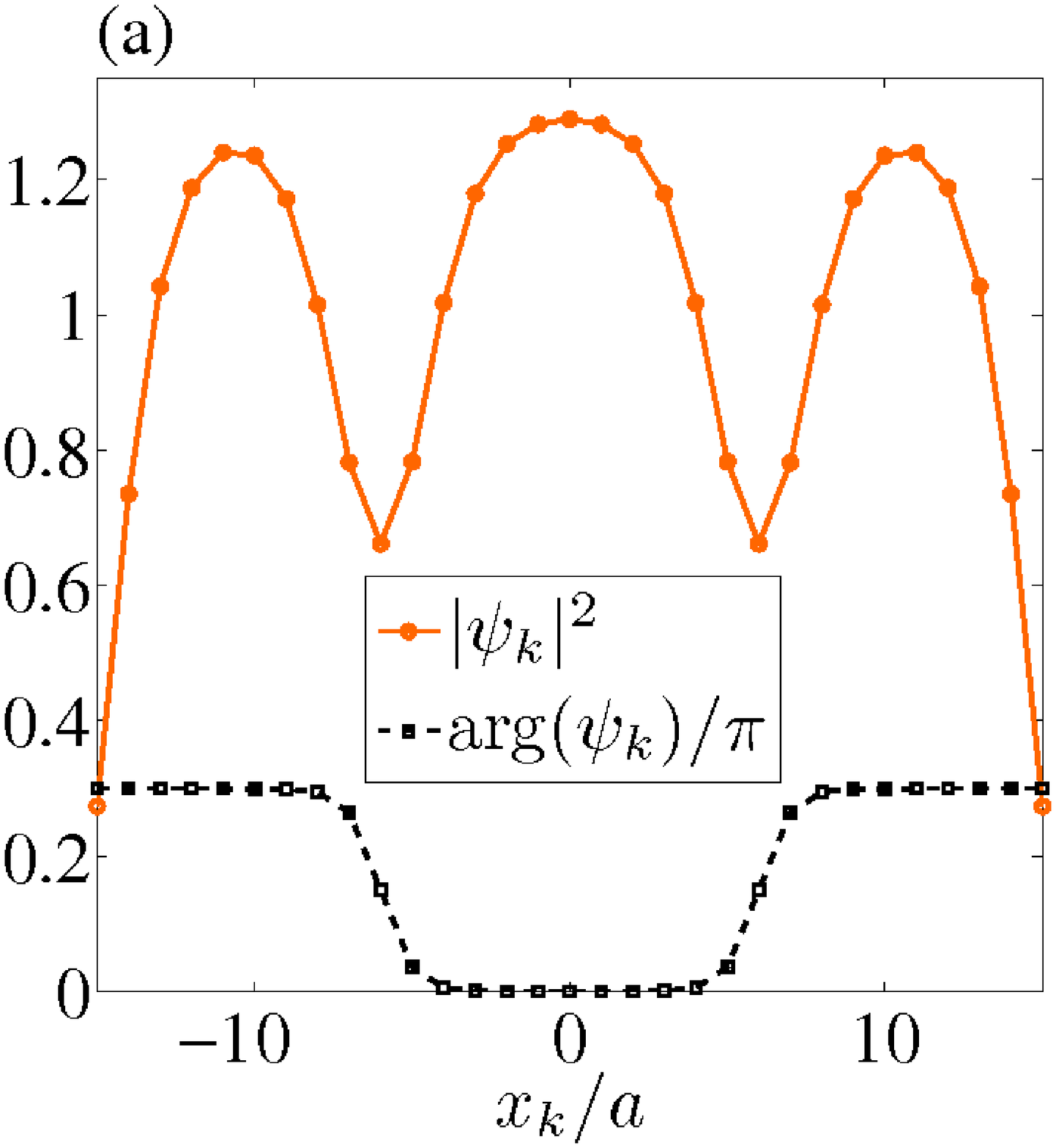}}}
\hspace{-0.0in}
\subfigure{\scalebox{\surfacePlotScale}{\includegraphics[angle=0]{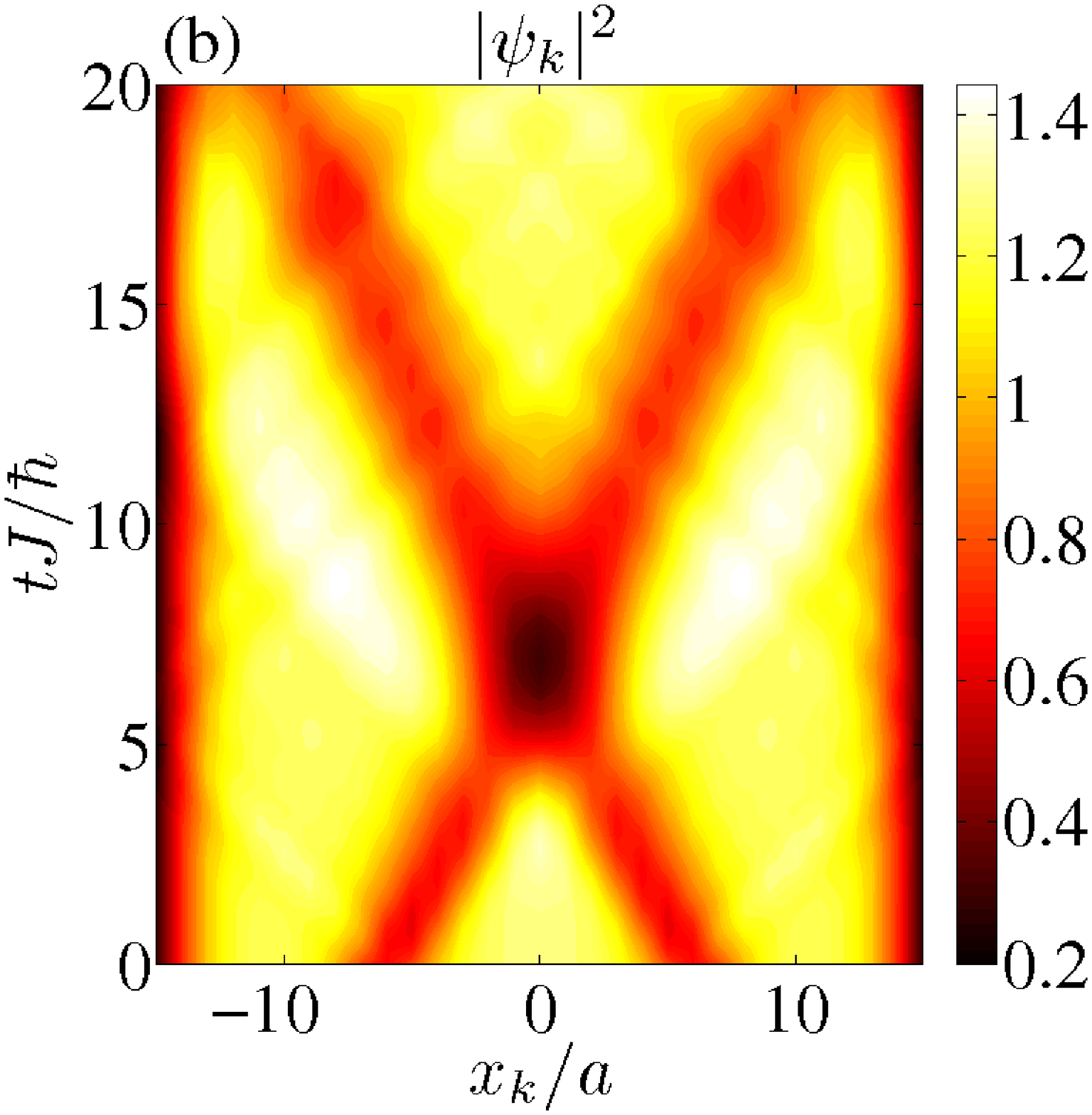}}}
\end{center}
\vspace{-0.25in}
\caption{(Color online) \emph{Engineered colliding soliton profile and DNLS dynamics.} (a) As in Fig.~\ref{fig:eng_standing_DNLS}, we show the DNLS initial condition, but this time we plot the state that results in a symmetric soliton collision during time evolution.  (b) The resulting dynamics according to the DNLS reveal an elastic collision between the two dark solitons. \label{fig:eng_colliding_DNLS}}
\end{figure}

One can apply the very same soliton engineering methods used above to create initial conditions in both the DNLS and BHH that result in a collision between two dark solitons traveling at equal-and-opposite velocities. In this case, we use a potential of the form
\begin{equation}
\epsilon_k = V_2\left\{\exp{\left[-\frac{(x_k+b)^2}{2 \sigma_{V2}^2}\right]}+
\exp{\left[-\frac{(x_k-b)^2}{2 \sigma_{V2}^2}\right]}\right\}
\label{eqn:DNLSgaussPot2}
\end{equation}
for the density engineering.  This potential has the effect of creating two identical density notches at positions $x=\pm b$ with depths and widths controlled by $V_2$ and $\sigma_{V2}$, respectively.  We then imprint an instantaneous phase of the form
\begin{align}
\theta_k = \Delta\theta_2
\{&-\frac{1}{2}\tanh\left[\frac{2(x_k+b)}{\sigma_{\theta 2}}\right] \nonumber \\ &+\frac{1}{2}\tanh\left[\frac{2(x_k-b)}{\sigma_{\theta 2}}\right]+1 \}.
\label{eqn:DNLSphaseImprint2}
\end{align}
The speed of the solitons is controlled by the phase drop $\Delta\theta_2$, and phonon generation is minimized by appropriately tuning the width $\sigma_{V2}$ of the phase profiles to the soliton depth as determined by $V_2$ in the density engineering stage.

In Fig.~\ref{fig:eng_colliding_DNLS}, we create a soliton-soliton collision in the DNLS.  The initial condition obtained after density and phase engineering is shown in Fig.~\ref{fig:eng_colliding_DNLS}(a), while the subsequent evolution of the density is shown in Fig.~\ref{fig:eng_colliding_DNLS}(b).  For parameters, we take $\nu U/J=0.35$ at filling $\nu=1$ with $M=30$ lattice sites and $V_2/J=0.4$, $\sigma_{V2}/a=1$, $b/a=6$, $\Delta\theta_2=0.3\,\pi$, and $\sigma_{\theta 2}/a=2$.  The collision is elastic, i.e., the speeds of the solitons are unchanged as a result of the collision.  The forward in time (backward in space) trajectory shifts of the solitons due to the collision are barely noticeable by eye~\cite{carr2002a}.  We show the results of the analogous TEBD simulation in Fig.~\ref{fig:eng_colliding_number}(a) with numerical parameters $\chi=80$ and $d=7$.  For this filling factor ($\nu=1$), the average particle number density $\langle\nhat_k\rangle$ closely follows the DNLS wave function density $|\psi_k|^2$, especially at times before and during the collision.

\begin{figure}
\begin{center}
\subfigure{\scalebox{\surfacePlotScale}{\includegraphics[angle=0]{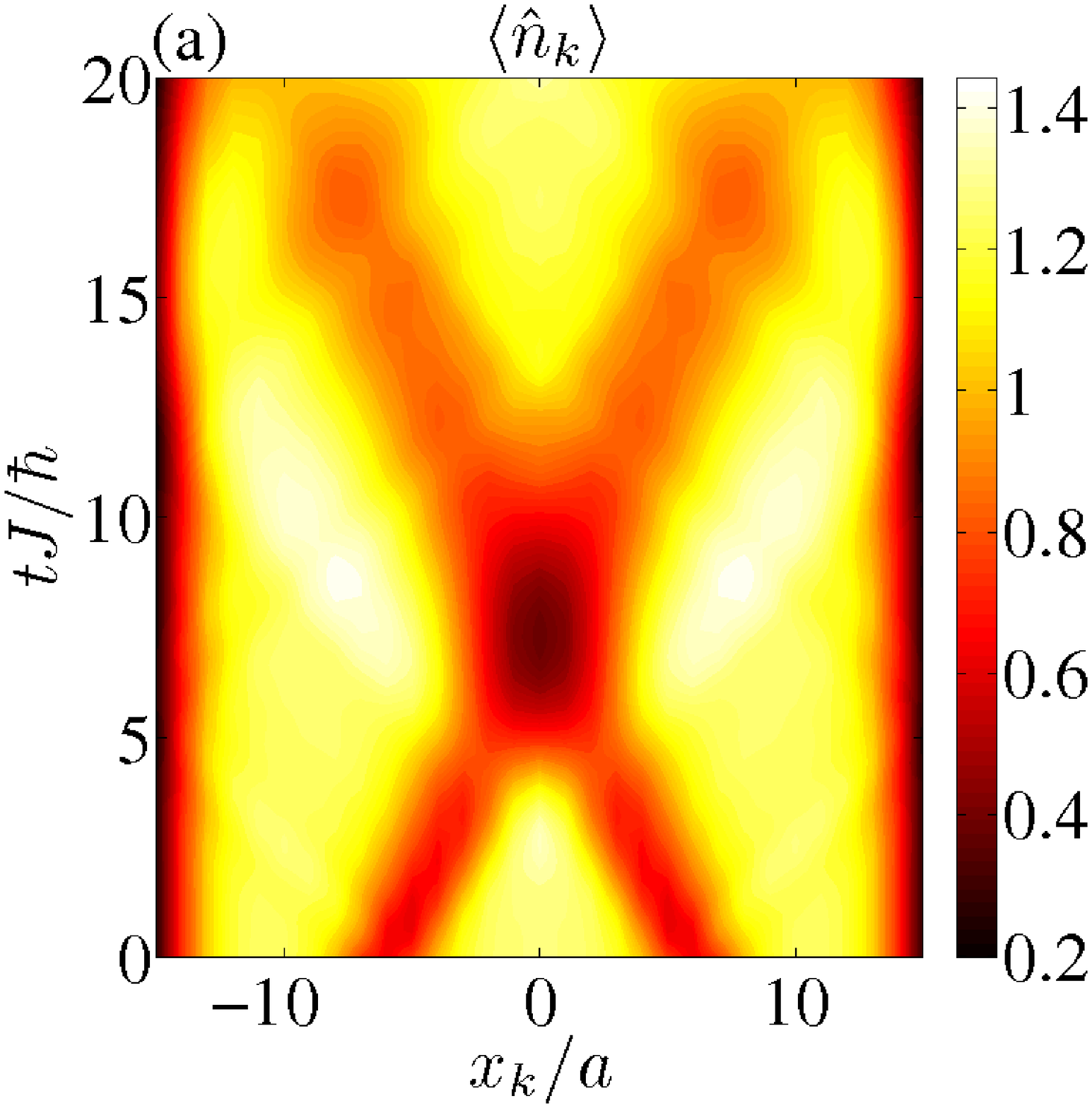}}}
\hspace{-0.0in}
\subfigure{\scalebox{\surfacePlotScale}{\includegraphics[angle=0]{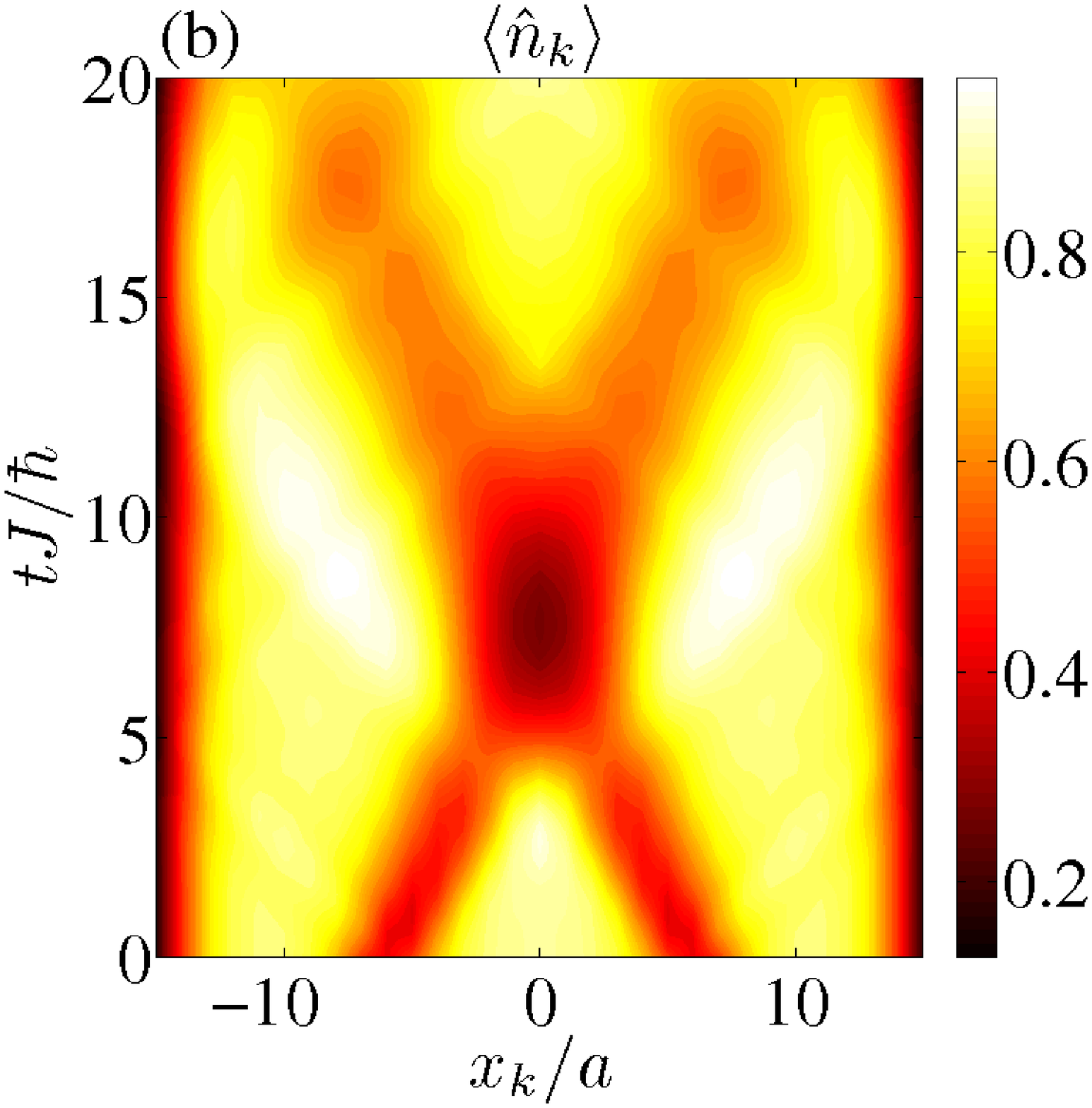}}}
\end{center}
\vspace{-0.30in}
\begin{center}
\subfigure{\scalebox{\surfacePlotScale}{\includegraphics[angle=0]{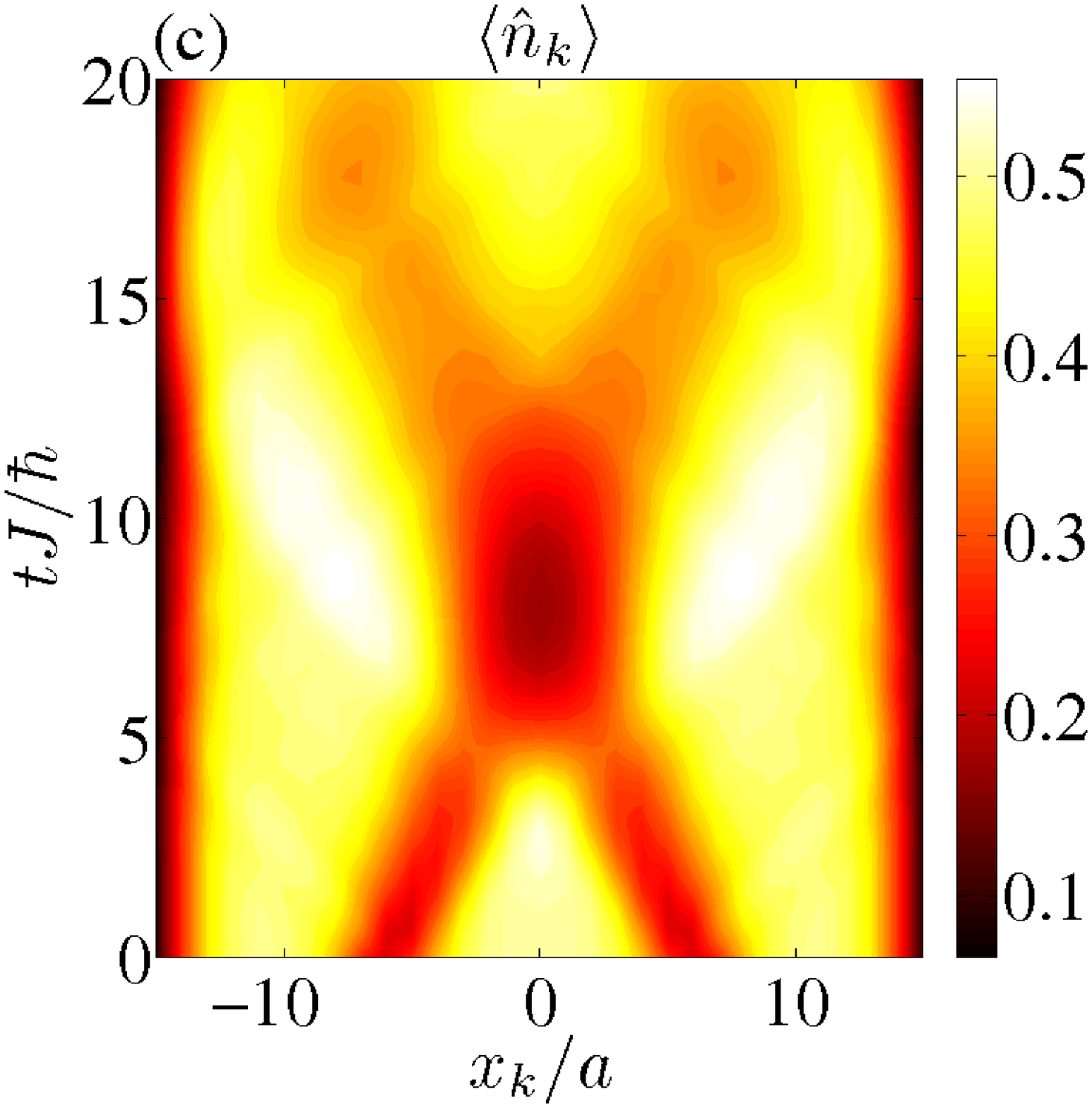}}}
\hspace{-0.0in}
\subfigure{\scalebox{\surfacePlotScale}{\includegraphics[angle=0]{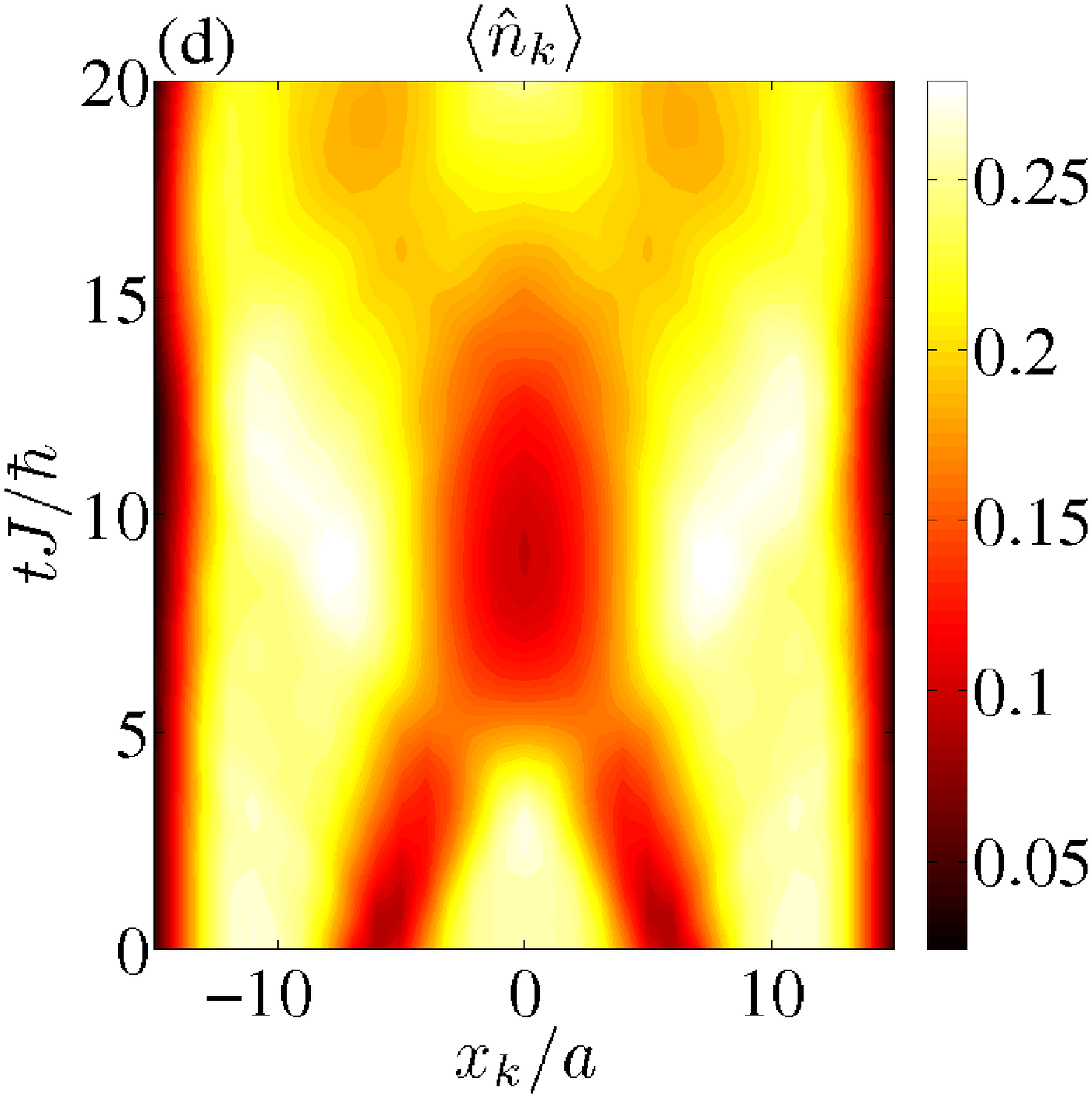}}}
\end{center}
\vspace{-0.25in}
\caption{(Color online) \emph{Quantum-induced inelasticity for quantum engineered dark solitons.} A collision between two engineered dark quantum solitons is shown for fixed $\nu U/J=0.35$ at filling factors (a) 1, (b) 0.7, (c) 0.4, and (d) 0.2.  The average particle number is plotted in each case.  The collision elasticity decreases with decreased filling due to an increase in the growth rate of quantum fluctuations for lower particle densities. \label{fig:eng_colliding_number}}
\end{figure}

However, as occurs for the standing soliton analyzed above, there exist two-body scattering processes which deplete the system and give rise to entangled quantum dynamics. The depletion processes which preserve the overall symmetry of the many-body wave function include two atoms scattering from a symmetric mode into an antisymmetric mode and two atoms scattering from a symmetric mode into another symmetric mode of higher order. Since antisymmetric modes must vanish at $x=0$, occupation of such modes will affect a collision of two solitons that occurs at the center of the lattice.  In order to see this effect, we can lower the filling factor $\nu$ while keeping $\nu U/J$ fixed in order to increase the rate of quantum fluctuations. Then there is a significant amount of depletion before or during the collision.  In panels (b), (c), and (d) of Fig.~\ref{fig:eng_colliding_number}, we show the quantum evolution with the same soliton engineering parameters and $\nu U/J=0.35$ but with $\nu=0.7, 0.4, 0.2$, respectively.  Because $\nu U/J$ is held fixed, the initial density is the same in each case up to a scale factor, but the different growth rates of depletion give rise to different density profiles at later times.  In fact, because the mode into which most atoms deplete is antisymmetric and vanishes at $x=0$, tuning the time scale of quantum fluctuations to occur near the time of collision effectively induces an inelasticity in the collision.  This can clearly be seen by comparing the $\nu=1$ case to the $\nu=0.2$ case in Fig.~\ref{fig:eng_colliding_number}.  We have observed qualitatively similar results when considering DNLS mean-field initial conditions for subsequent BHH quantum evolution of a pair of colliding solitons~\cite{Mishmash09_short, Mishmash08_IMACS}.

This result presents another strong piece of evidence that a given experiment will not reveal the solitons drifting to the left or right randomly without changing shape, i.e., the solitons are not simply spreading out or diffusing without dissipating:  if the solitons were spreading out, they would not appear to stick together on average over many density measurements as in Fig.~\ref{fig:eng_colliding_number}.  Hence, the phenomenon of quantum-induced inelasticity complements the result regarding filling-in of the pair correlation function in the standing soliton case [see Fig.~\ref{fig:eng_standing_quantum}(a)]:  when a fully quantum many-body simulation of the dark soliton is performed, it predicts that the soliton gets depleted and fills in, even during a \emph{single} experiment.

\section{Quantum Evolution of Mean-Field Solitons} \label{sec:DNLSsolitons}

In the above analysis, we used density and phase engineering to generate initial conditions in the BHH that resemble dark solitons.  However, the dark soliton is the first excited stationary state of the DNLS.  In this section, we consider what happens when such a mean-field soliton state is injected directly into the quantum BHH.  Two of us have analyzed this case in two other works as well~\cite{Mishmash08_IMACS, Mishmash09_short}, so we will only treat it briefly.  Here we present new results in terms of Pegg-Barnett phase averages, as compared to previous work.

To summarize the method, we first perform constrained imaginary time relaxation on the DNLS with an initial condition in the form of a linear function through the center of the lattice. We run the imaginary time evolution until converged to the fundamental dark soliton solution of the DNLS with a notch centered at $x=0$.  We then use a product of on-site coherent states, cf. Eq. (\ref{eqn:cohState1}), truncated at $d$ number states to build a Fock space representation of the DNLS soliton solution.  The matrix product state representation of such a product state is trivial, so performing subsequent evolution in TEBD is straightforward~\cite{Mishmash08_IMACS}.  It is important to note that in this case we must use a non-number-conserving TEBD routine because the initial mean-field states themselves, i.e., truncated coherent states, do not conserve total particle number due to DNLS mean-field theory assuming that the U(1) symmetry present in the BHH is spontaneously broken.

\begin{figure}
\begin{center}
\subfigure{\scalebox{\surfacePlotScale}{\includegraphics[angle=0]{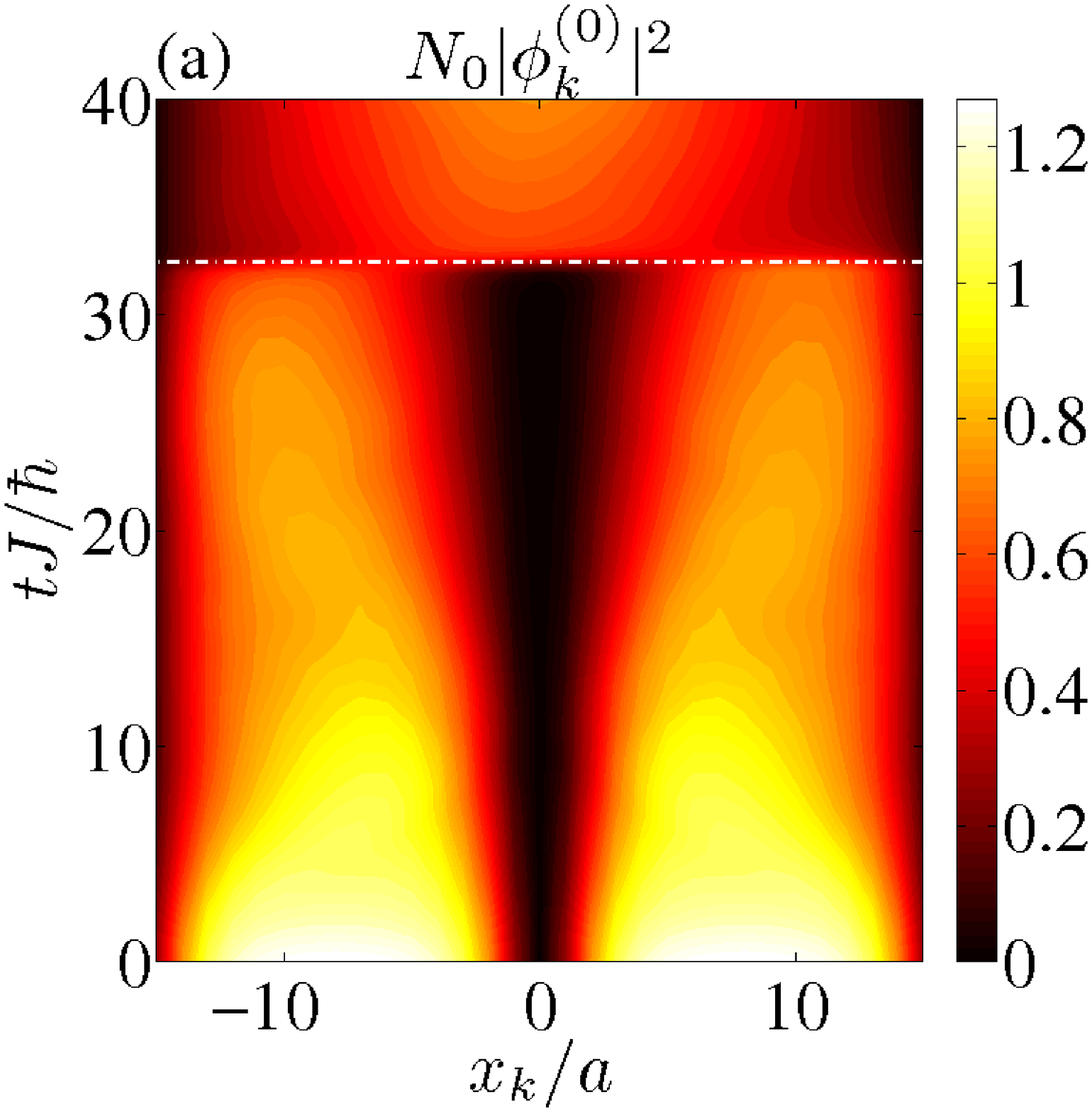}}}
\hspace{-0.0in} \subfigure{\scalebox{\surfacePlotScale}{\includegraphics[angle=0]{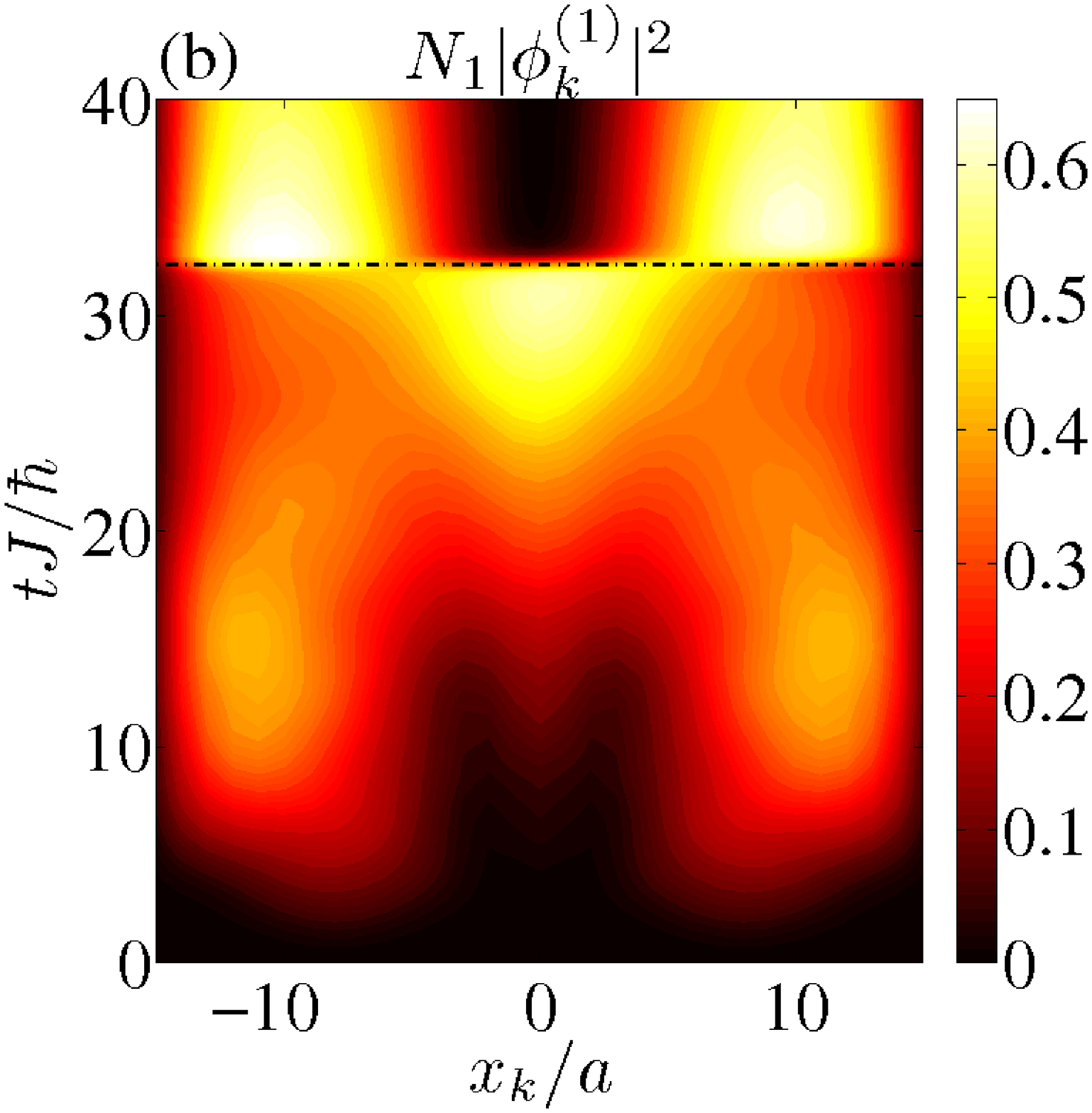}}}
\end{center}
\vspace{-0.3in}
\begin{center}
\subfigure{\scalebox{\surfacePlotScale}{\includegraphics[angle=0]{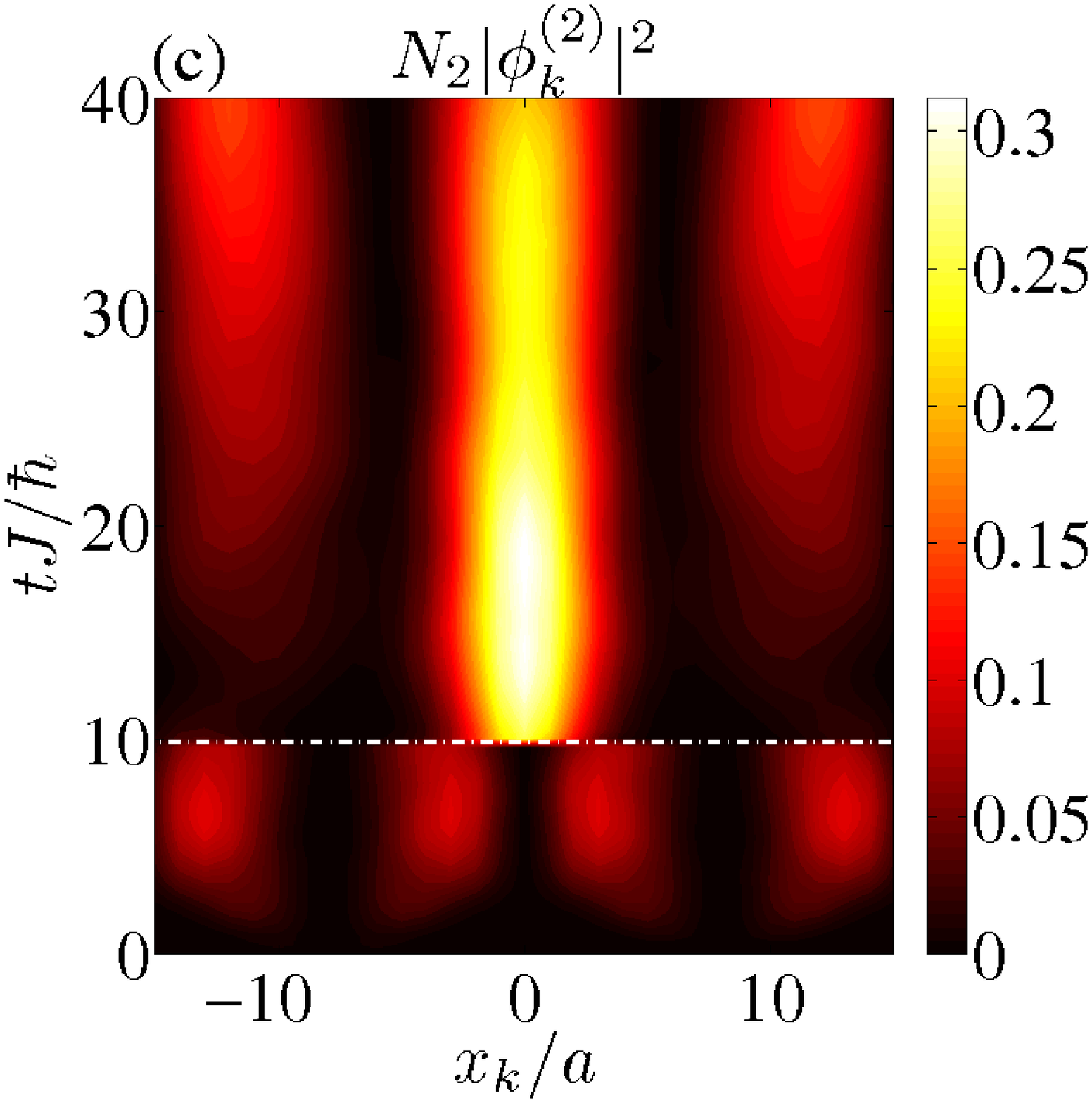}}}
\hspace{-0.0in} \subfigure{\scalebox{\surfacePlotScale}{\includegraphics[angle=0]{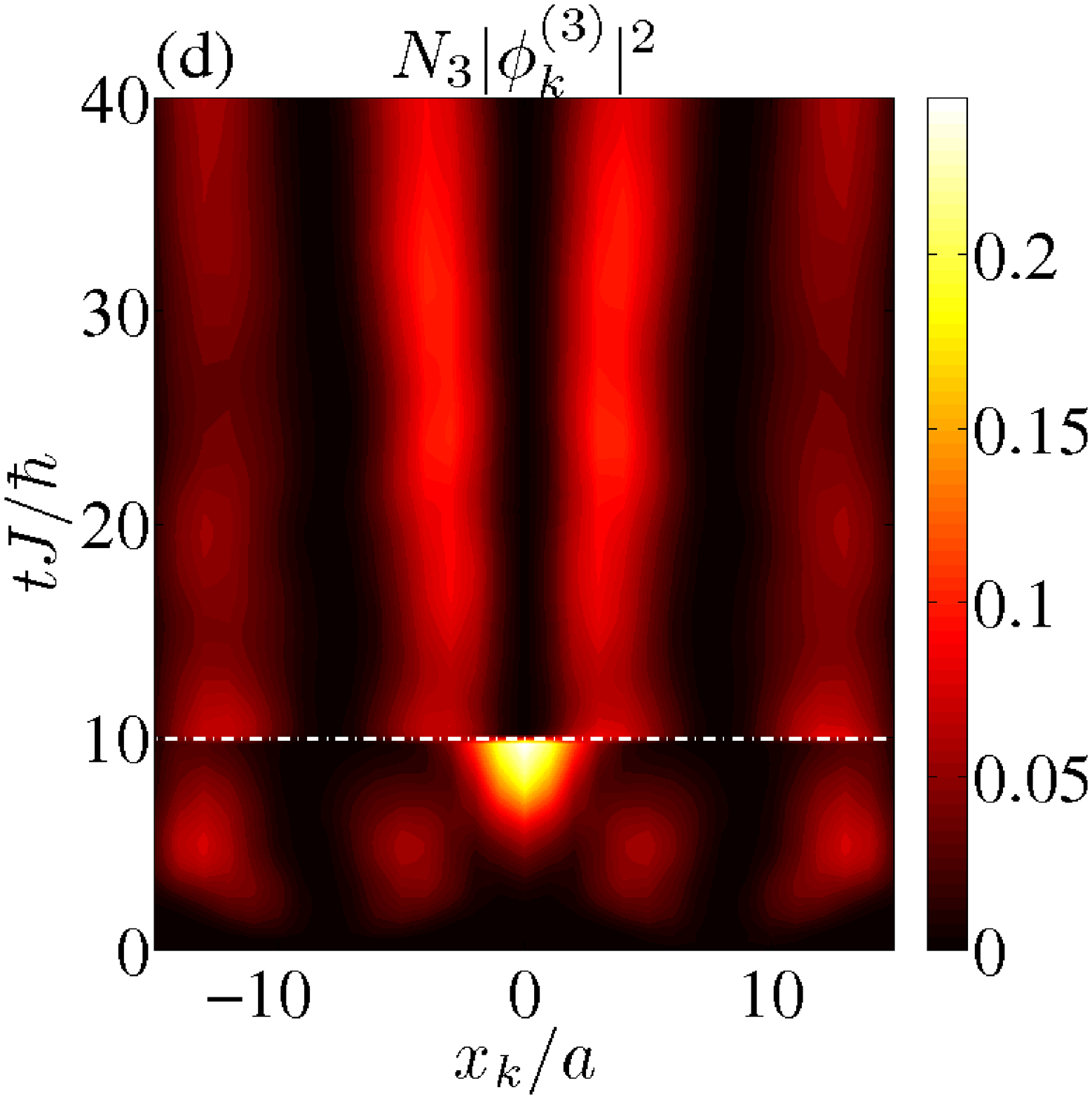}}}
\end{center}
\vspace{-0.25in}
\caption{(Color online) \emph{Natural orbital dynamics during quantum evolution of a standing mean-field soliton.}  (a) The condensate wave function and the (b) second, (c) third, and (d) fourth most highly occupied natural orbitals during quantum evolution of a mean-field dark soliton.  Note the different color bars in each case.  The discontinuities are due to a swapping of natural orbital occupation numbers at specific times during the evolution as we indicate with dashed-dotted horizontal lines. \label{fig:DNLS_standing_natorbs}}
\end{figure}

In this case, as in Sec.~\ref{sec:QuantSoliEng}, we observe the initial soliton configuration to fill in with depleted atoms over time~\cite{Mishmash09_short}.  The analysis of the depletion mechanism is very similar to that presented above.  In Fig.~\ref{fig:DNLS_standing_natorbs}, we show the quantum dynamics of the four most highly occupied natural orbitals, cf. Fig.~\ref{fig:eng_standing_natorbs} for the analogous plot for density and phase engineered initial conditions, using TEBD with parameters $\nu U/J=0.35$, $\nu=1$, $M=31$, $\chi=50$, $d=7$.  The quantity $\nu=N_\mathrm{DNLS}/M$ is now the \emph{average} filling factor with $N_\mathrm{DNLS}$ being the norm of the initial DNLS soliton wave function.  We will refer to the plots in Fig. \ref{fig:DNLS_standing_natorbs} when conducting a BDG analysis of the initial DNLS dark soliton state in Sec. \ref{sec:Bog}.

\begin{figure}
\begin{center}
\subfigure{\scalebox{\surfacePlotScale}{\includegraphics[angle=0]{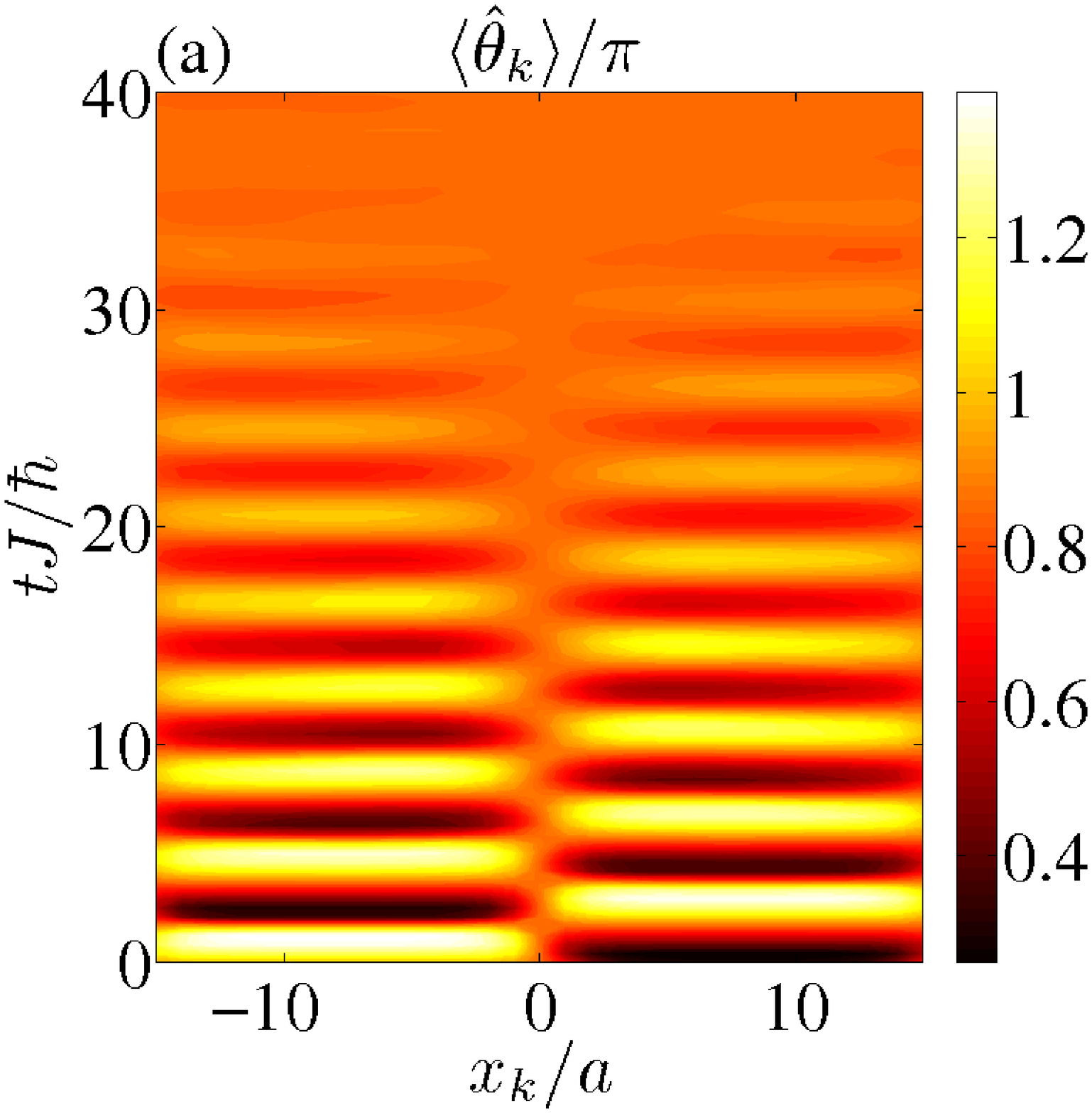}}}
\hspace{-0.0in} \subfigure{\scalebox{\surfacePlotScale}{\includegraphics[angle=0]{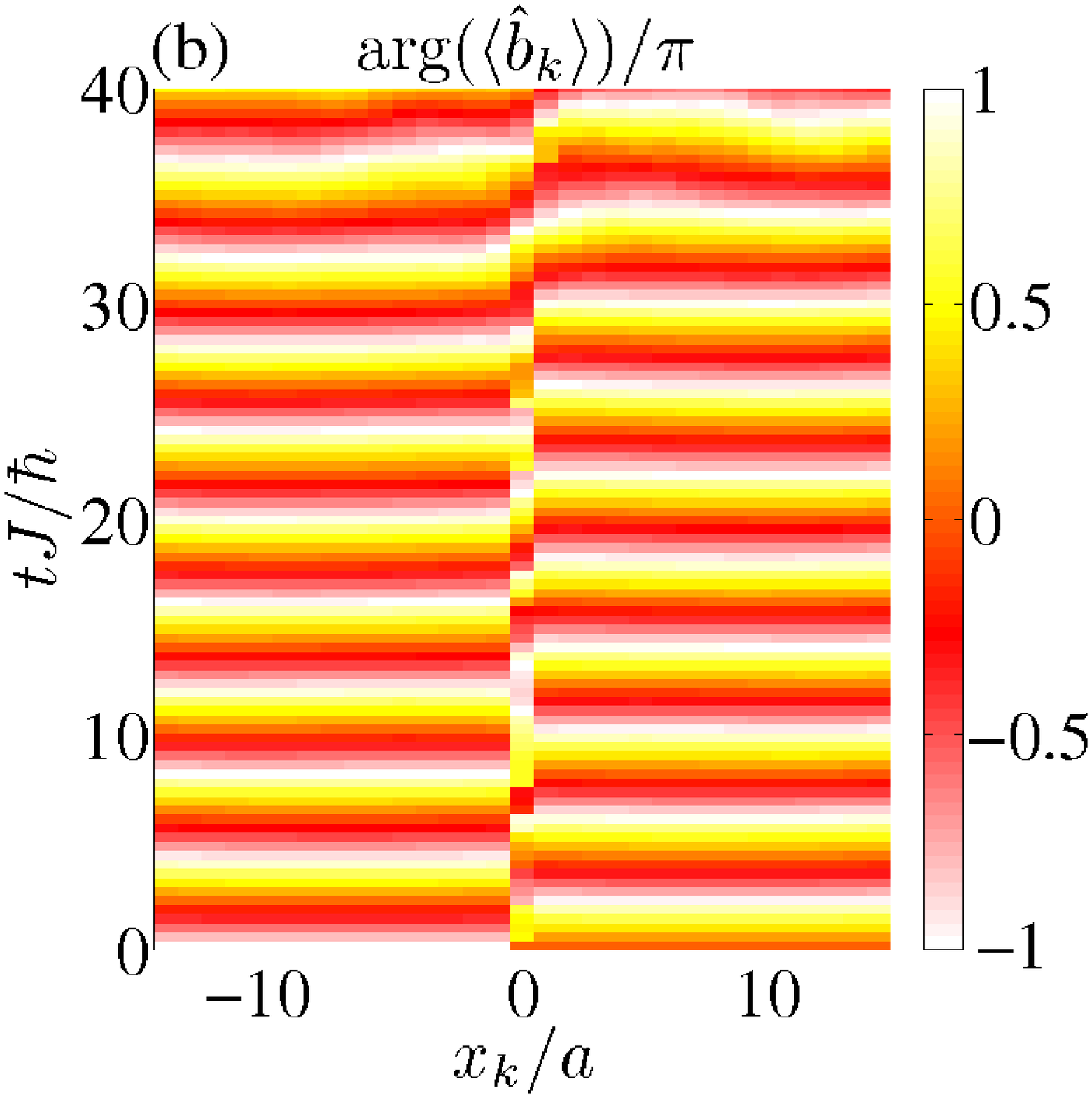}}}
\end{center}
\vspace{-0.25in}
\caption{(Color online) \emph{Phase measures during quantum evolution of a standing mean-field soliton.}  (a)  The average of the Pegg-Barnett phase operator and (b) the phase of the order parameter during quantum evolution of a mean-field dark soliton.  The phase drop across the notch according to the Pegg-Barnett definition decays as the soliton depletes, whereas the order parameter phase maintains a $\pi$ phase drop across $x=0$ for all times.  \label{fig:DNLS_standing_phase}}
\end{figure}

Of course, a density notch is not all that characterizes a dark soliton; the phase drop across the notch is equally important.  In fact, in the context of the mean-field GP equation, it is exactly this phase drop that stabilizes a stationary dark soliton and keeps particles from filling it in.  Thus, loosely speaking, we can say that quantum many-body effects are ``stronger'' than this phase drop in that these effects do cause the soliton to dissipate, as we have well shown.  In Fig.~\ref{fig:DNLS_standing_phase}, we consider two different measures of the phase of the soliton during quantum evolution:  (1) the average of the Pegg-Barnett phase operator, $\langle\thetahat_k\rangle$, and (2) the argument of the order parameter, $\arg(\langle\bhat_k\rangle)$.  We can also calculate the phase of the natural orbitals, i.e., $\arg(\phi_k^{(j)})$.  In fact, the symmetries of the natural orbitals were alluded to in the discussion of the soliton depletion mechanisms in Sec.~\ref{sec:QuantSoliEng}, but as mentioned there, it is not insightful to plot the quantities $\arg(\phi_k^{(j)})$ over time.

We see in Fig.~\ref{fig:DNLS_standing_phase}(a) that at time $t=0$ the Pegg-Barnett phase operator initially exhibits a phase drop across the notch approximately equal to $\pi$.  However, when particles deplete out of the antisymmetric soliton mode into modes of both even and odd symmetry, the phase drop across the notch according to the Pegg-Barnett definition decreases.  For times $t>0$ when multiple single-particle orbitals are occupied, the phase of the system in the single-particle wave function sense is ill-defined even though the Pegg-Barnett definition remains valid.  Of course, $\thetahat_k$ being a local observable, this is an effect that manifests itself only over many measurements.  It is also important to state that the on-site Pegg-Barnett phase operator as we have defined it requires a breaking of the U(1) symmetry, i.e., non-conservation of particle number:  the only nontrivial part (the second term) of $\thetahat_k$ in Eq. (\ref{eqn:phaseop}) will clearly vanish when averaged over a many-body state with conserved particle number.

In Fig.~\ref{fig:DNLS_standing_phase}(b), we show the space-time dependence of the phase of the on-site order parameter.  As the soliton depletes, the norm of the order parameter, $|\langle\bhat_k\rangle|^2$, decays in the tails of the soliton~\cite{Mishmash09_short}.  However, as clearly shown in Fig.~\ref{fig:DNLS_standing_phase}(b), the $\pi$ phase drop of the order parameter across the notch persists in time.  Next, we seek to analyze the initial DNLS soliton configurations considered in this section within the BDG framework.

\section{Bogoliubov Analysis} \label{sec:Bog}

In this section, we use the Bogoliubov prescription to perform a linear stability analysis of the stationary state DNLS soliton solutions whose quantum dynamics were analyzed in Sec.~\ref{sec:DNLSsolitons} and Ref.~\cite{Mishmash09_short}.  We make an explicit comparison of the spectrum of elementary excitations predicted by the Bogoliubov theory versus the depletion of the soliton over time as predicted by quasiexact simulations of the BHH using TEBD.

Let us assume a stationary state of the DNLS so that the amplitude at each site $k$ evolves according to $\xi_k e^{-i\mu t/\hbar}$, where $\mu$
is the associated chemical potential.  We then perturb this initial condition and look for solutions to the DNLS of the form
\begin{equation}
\psi_k(t) = \left\{\xi_k + \sum_j\left[u_k^{(j)} e^{-i\omega_j t} + v_k^{(j)\ast} e^{i\omega_j^\ast t}\right] \right\}e^{-i\mu t/\hbar}, \label{eqn:BogAnsatz}
\end{equation}
where $u_k^{(j)}$ and $v_k^{(j)}$ are the amplitudes of the $j^{\mathrm{th}}$ normal mode
and $\omega_j$ is the associated mode frequency.  Insertion of Eq. (\ref{eqn:BogAnsatz}) into the DNLS results in the tight-binding form of the BDG equations which we solve numerically.  The diagonalization routine used to solve the BDG equations results in $2M$ eigenvectors and eigenvalues.  The norm of a given mode $j$ is defined as $\sum_{k=1}^M [|u_k^{(j)}|^2-|v_k^{(j)}|^2]$; this motivates the definition of the local mode density at site $k$ to be $|u_k^{(j)}|^2-|v_k^{(j)}|^2$.  For modes with nonzero positive norm, we renormalize the amplitudes such that $\sum_{k=1}^M[|u_k^{(j)}|^2- |v_k^{(j)}|^2]=1$.

If for a given mode the frequency $\omega_j$ is complex, the DNLS solution $\xi_k$ is said to be \emph{dynamically unstable}.  On the other hand, if $\omega_j$ is negative and the norm of the mode is positive, the mode is \emph{anomalous}, in which case the DNLS solution is at a local maximum in the energy landscape and is said to be \emph{energetically unstable}. A mode with identically zero frequency corresponds to a Goldstone mode associated with the breaking of a continuous symmetry in the stationary mean-field solution.  As an example, we always observe in our numerical BDG calculations the Goldstone mode associated with the calculated DNLS soliton solution attaining an arbitrary global U(1) phase.

In a harmonic trap in the Thomas-Fermi limit, the dark soliton has an anomalous mode with a frequency $-1/\sqrt{2}$ times the trapping frequency~\cite{Busch00_PRL_84_2298, Dziarmaga02_PRA_66_043620}; this mode is associated with translation of the soliton in the trap.  In a system with full translational symmetry, this mode is a Goldstone mode and thus has zero frequency.  In the continuum, such a zero frequency (and zero norm) translational mode also appears in a box geometry given that the healing length is much smaller than the box size and the soliton is placed sufficiently far from the hard walls~\cite{Dziarmaga04_PRA_70_063616}.  In our case, we should not expect this mode to have exactly zero frequency since the lattice itself breaks continuous translational symmetry.  Under these guidelines, we now proceed to analyze the solutions of the BDG equations for a stationary DNLS soliton background.

\begin{figure}
\begin{center}
\subfigure{\scalebox{0.19}{\includegraphics[angle=0]{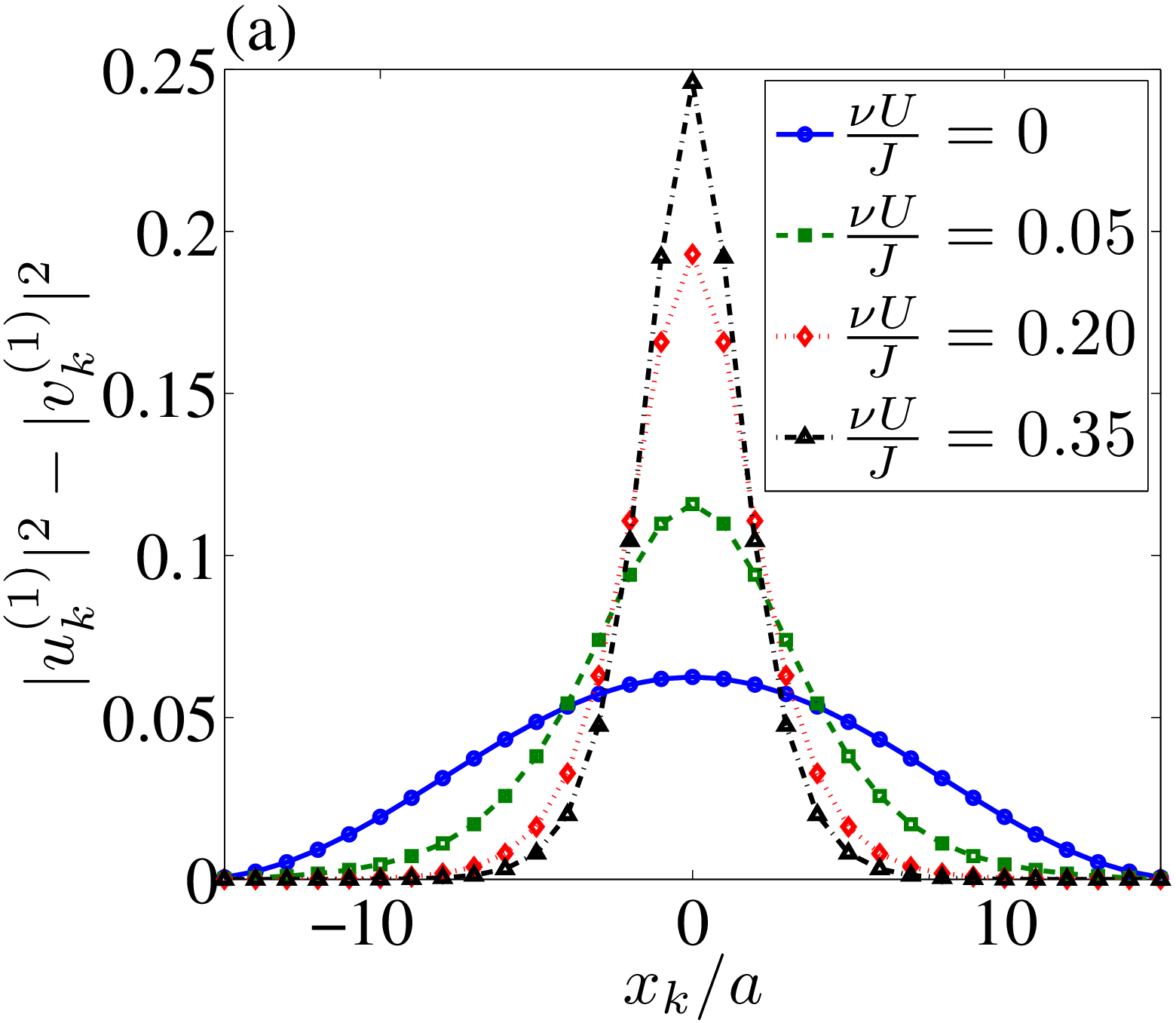}}}
\hspace{-0.1in}\subfigure{\scalebox{0.19}{\includegraphics[angle=0]{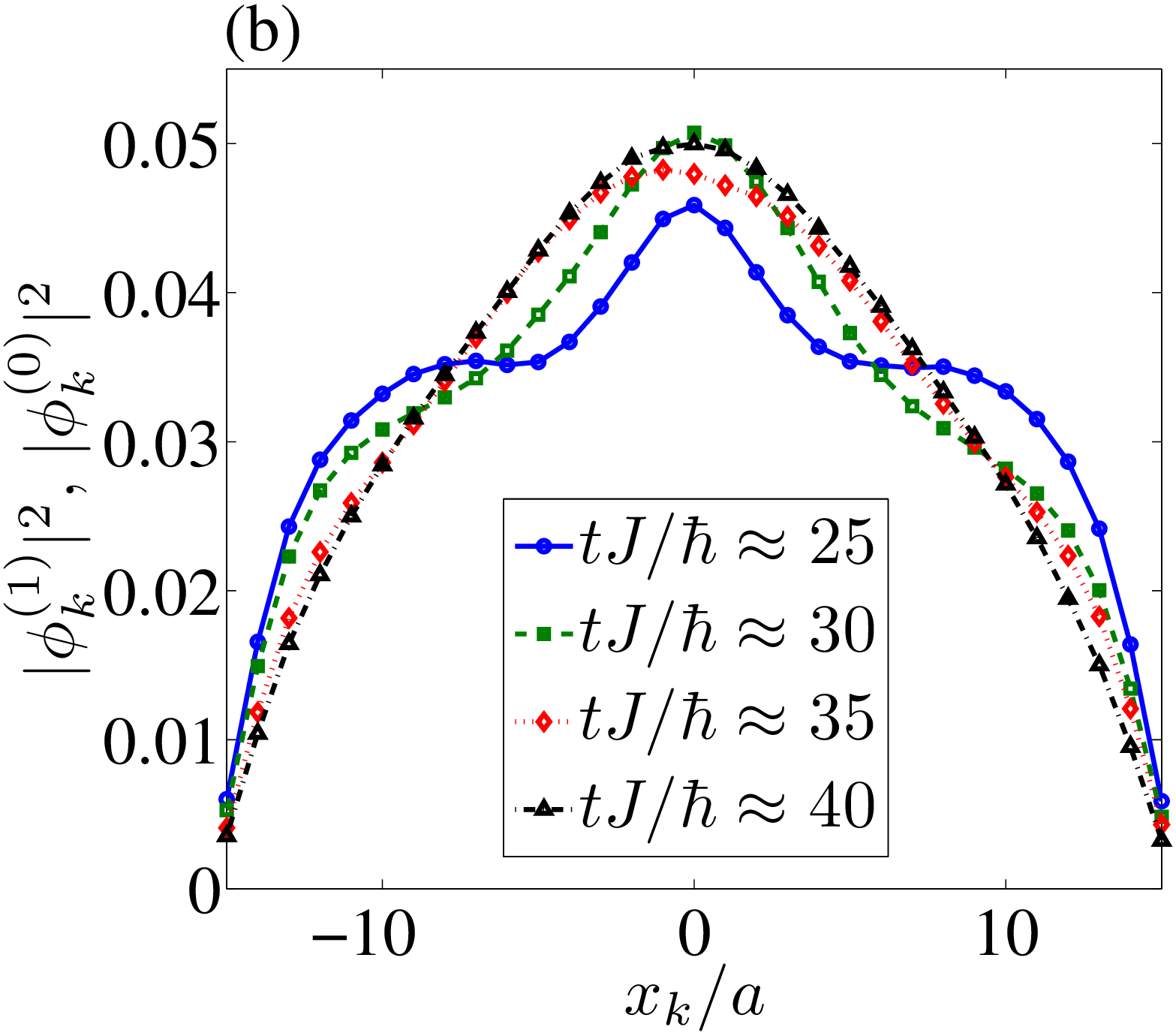}}}
\end{center}
\vspace{-0.20in} \caption{(Color online) \emph{Anomalous Bogoliubov mode:  Density profile and comparison to first depleted mode.} (a) The density of the anomalous (negative frequency and positive norm) mode of the Bogoliubov excitation spectrum is plotted versus space for effective interactions $\nu
U/J=0,0.05,0.20,0.35$ at $M=31$ sites. (b) For the latter case of $\nu U/J$, we show the time
evolution at time steps $tJ/\hbar\approx 25,30,35,40$ of the depleted natural
orbital most responsible for the decay of the dark soliton during the initial stages of evolution.  We see that the Bogoliubov mode in (a)
for $\nu U/J=0.35$ does not resemble the evolving symmetric mode shown in (b).
\label{fig:DNLS_standing_BogAnomModes}}
\end{figure}

For the case of $M=31$ sites as in Sec.~\ref{sec:DNLSsolitons} and Ref.~\cite{Mishmash09_short}, for interaction strengths of interest, $\nu U/J\lesssim0.50$, we observe in our calculations an anomalous mode with small negative frequency $-0.05 J/\hbar\lesssim\omega_1<0$.  As expected, this anomalous mode is characterized by a density maxima concentrated in the region of the soliton notch as shown in Fig.~\ref{fig:DNLS_standing_BogAnomModes}(a) for various values of $\nu U/J$.  We wish to compare the density of this anomalous Bogoliubov mode to the most highly occupied symmetric natural orbital during full quantum evolution.  Figure~\ref{fig:DNLS_standing_BogAnomModes}(b) depicts the density of the symmetric natural orbital with highest occupation at times $tJ/\hbar\approx 25,30,35,40$ during quantum evolution according to the BHH of the stationary DNLS soliton [see Fig.~\ref{fig:DNLS_standing_natorbs}(a)--(b)].  Strictly speaking, this mode is $\phi_k^{(1)}$ for $tJ/\hbar\lesssim 33$ and $\phi_k^{(0)}$ for $tJ/\hbar\gtrsim 33$ due to the ordering of the natural orbitals being defined to coincide with the ordering of occupation numbers.  Recall that for this simulation $\nu U/J=0.35$.  It is clear that the symmetric natural orbital in Fig.~\ref{fig:DNLS_standing_BogAnomModes}(b), the natural orbital most responsible for depletion out of the soliton orbital, does not closely resemble the negative frequency mode of the Bogoliubov theory:  this natural orbital, unlike the anomalous Bogoliubov mode has significant density away from the lattice center.  Also, as can be seen in Fig.~\ref{fig:DNLS_standing_natorbs}, there is also significant contribution from natural orbitals of higher order, e.g., $\phi_k^{(2)}$, $\phi_k^{(3)}$, etc., to the filling-in effect.  A truncation of the Bogoliubov Hamiltonian at the anomalous mode would thus hardly be justified in this regime of strong quantum fluctuations.

\begin{figure}
\begin{center}
\hspace{-0.0in}\subfigure{\scalebox{0.35}{\includegraphics[angle=0]{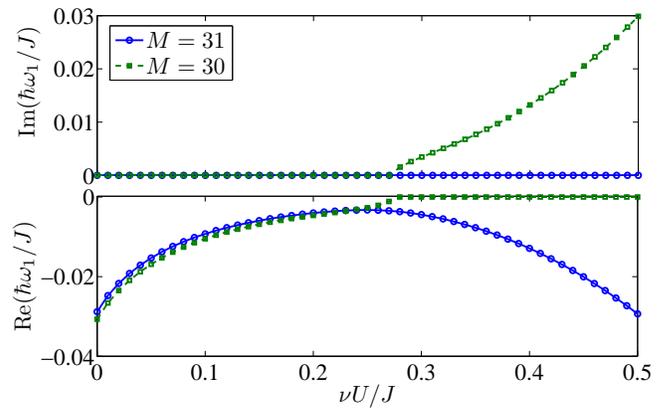}}}
\end{center}
\vspace{-0.20in} \caption{(Color online) \emph{Frequency of Bogoliubov mode versus interaction strength:  Appearance of energetic and dynamical instability.}
For $M=31$ and $M=30$ sites, we show the real and imaginary parts of the calculated negative or complex frequencies appearing in the Bogoliubov excitation spectrum.  For an odd number of sites, e.g., $M=31$, the frequency is always negative and real, while for an even number of sites, e.g., $M=30$, the frequency becomes purely imaginary past a critical value of $\nu U/J$.  This behavior is general for any odd or even number of sites and is not specific to the particular choice of lattice sizes made above. \label{fig:DNLS_standing_BogFreqs}}
\end{figure}

The dependence of the frequency of the anomalous Bogoliubov mode versus effective interaction strength $\nu U/J$ for the case of $M=31$ sites is depicted as the solid blue curve connecting circle points in Fig.~\ref{fig:DNLS_standing_BogFreqs}.  We can understand the form of this dependence as follows.  The anomalous mode is associated with the translation of the soliton in our lattice plus box geometry and in general could have nonzero negative frequency due to the lattice and box breaking full translational symmetry.  For the case of $\nu U/J=0$, our ``soliton'' state is the first excited state of the linear tight-binding \schrod equation.  The size of the density notch is thus on the order of the system size and the box works to break the translational symmetry associated with the density notch position in a manner similar to how a harmonic trap breaks this symmetry:  for $\nu U/J=0$, the translational mode is always anomalous. As we increase $\nu U/J$, we are decreasing the healing length, i.e., the soliton size, and thus restoring translational symmetry of the soliton \emph{inside the box}; thus, the frequency of the anomalous mode becomes less negative.  In the continuum, the mode would eventually become a zero mode, cf. Ref.~\cite{Dziarmaga04_PRA_70_063616}.  However, as we continue to increase $\nu U/J$ and decrease the soliton size, the breaking of translational symmetry \emph{by the lattice} begins to play a role.  Past some critical value of $\nu U/J$ at which $\omega_1$ is nearly zero, the frequency goes more negative upon further increasing $\nu U/J$.  This picture is general for the case of an odd number of sites in which the soliton notch is located at a particular site.  Recall that for simplicity, we are here only considering solitons placed at the lattice center.  Also, as expected, the critical value of $\nu U/J$,  past which $\omega_1$ begins to decrease, decreases as the system size $M$ is increased.  This is because the soliton size goes like $\sqrt{J/\nu U}$, and the larger the lattice size is, the more quickly translational symmetry within the box is restored as $\nu U/J$ is increased from zero.

On the other hand, for the case of an even number of sites in which the notch of the solitonic stationary state of the DNLS is located between sites, the behavior is qualitatively different from that described above for an odd number of sites. As shown in Fig. \ref{fig:DNLS_standing_BogFreqs} for $M=30$, the negative frequency mode goes to zero at some critical value of $\nu U/J$.  Now, rather than becoming more negative past this critical value as for $M=31$, the mode instead becomes complex, thus indicating the onset of a discreteness-induced dynamical instability.
The emergence of such a dynamical instability has been pointed out also for a dark soliton state
in a double-well potential~\cite{Raghavan99_PRA_59_620, Ichihara08_PRA_78_063604}, which corresponds to the limit of the small even number of sites, $M=2$.
However, we have checked that quantum evolution of DNLS dark solitons for an even number of sites is qualitatively identical to the behavior depicted in Sec.~\ref{sec:DNLSsolitons} and Ref.~\cite{Mishmash09_short} in which an odd number of sites was assumed.  We could also use the imaginary part of the complex frequency mode to estimate the timescale over which dynamical instability would manifest itself for the soliton engineering calculations in Sec. \ref{sec:QuantSoliEng}, which for practical reasons were performed with an even number of sites.  Using the data in Fig. \ref{fig:DNLS_standing_BogFreqs}, we estimate that the timescale of dynamical instability for $\nu U/J=0.35$ and $M=30$ sites is three to four times longer than the longest times shown in the characteristic simulation of Sec. \ref{sec:QuantSoliEng}.

We have also considered higher order Bogoliubov modes, e.g., $(u_k^{(2)},v_k^{(2)})$, $(u_k^{(3)},v_k^{(3)})$, etc., for both the soliton and DNLS ground state background configurations.  According to our calculations, the lowest natural orbitals obtained using TEBD do not match the low energy Bogoliubov modes of the formed dark soliton state, although the second and third depleted natural orbitals, i.e., $\phi^{(2)}_k$ and $\phi^{(3)}_k$, compare more favorably to the second and third lowest energy Bogoliubov modes calculated with respect to the DNLS \emph{ground state}~\cite{Mishmash08_MSThesis}.  Still, as in Fig.~\ref{fig:DNLS_standing_BogAnomModes}(a)--(b), direct comparison is made difficult because the natural orbitals themselves are time-dependent objects.  Exactly how the noncondensed atoms distribute themselves among the single-particle natural orbitals cannot be accurately predicted by calculating the Bogoliubov quasiparticle spectrum of DNLS stationary states.  Finally, such a static Bogoliubov analysis as we have presented in this section cannot strictly be applied to the engineered dark solitons of Sec.~\ref{sec:QuantSoliEng} because the density and phase engineered initial conditions encountered in that case are not stationary states of even the DNLS.  It would in principle be possible to carry out time-dependent BDG calculations in the spirit of Ref.~\cite{Law03_PRA_68_015602}, treating the soliton engineering as part of the dynamical process.  However, the substantial amount of quantum depletion and quantum entanglement present as predicted by our full quantum TEBD simulations suggests that such a theory based on noninteracting Bogoliubov quasiparticles would inevitably be inapplicable.  On a related note, breakdown of BDG theory when applied to ground state properties of the 1D BHH is known to occur for interaction strengths as low as $U/J\gtrsim 0.2$~\cite{Danshita09_PRA_79_043601}.

\section{Conclusions} \label{sec:ThatsAll}

We have shown that dark solitons decay in the presence of strong quantum fluctuations induced by an optical lattice, even in the superfluid regime of the Bose-Hubbard Hamiltonian.  Our work is complementary to past mainly perturbative studies of quantum fluctuations in which it was found that dark solitons delocalize but do not decay.

Specifically, we used time-evolving block decimation to first engineer one or more dark solitons in an experimentally accessible manner through manipulation of the lattice potential itself.  We then followed the entangled dynamics of this decay by evaluating a large suite of quantum measures, including several entanglement measures, moments of number and Pegg-Barnett phase operators, quantum depletion, and second-order correlations in the form of $g^{(2)}$.  We showed that not only the mean field but also the Boguliubov-de-Gennes equations break down, even in the superfluid regime of the phase diagram.  We made an explicit comparison of BDG analysis and our entangled dynamical simulations, and found that ``depletion modes,'' i.e., higher natural orbitals in our more complete theory, do not in general resemble the BDG modes.

A main aim of this paper is to make previous work presented in Ref.~\cite{Mishmash09_short} relevant to realistic near-term experiments, as well as to compare our work in a more explicit way to past BDG-based studies~\cite{Dziarmaga03_JPhysB_36_1217, Dziarmaga06_JPhysB_39_57, Dziarmaga04_PRA_70_063616}.  The density and phase-engineering simulations of Sec.~\ref{sec:QuantSoliEng} can be realized in the lab with experimental parameters as follows.  The Bose-Hubbard effective interaction parameter $\nu U/J=0.35$ at filling $\nu=1$ corresponds to \Rb~atoms with an $s$-wave scattering length tuned via Feshbach resonance to $a_s=1.0$ nm in an optical lattice with longitudinal and transverse lattice heights $V_0=E_R$ and $V_{0\perp}=25\,E_R$ created with lasers of wavelength $\lambda=850~\mathrm{nm}$, where $E_R$ is the recoil energy of the lattice.  The Gaussian beam used for density engineering then has height $V_1=J\approx0.17\,E_R$ and width $\sigma_{V1}=a=425~\mathrm{nm}$. The width of the fall-off of the tanh phase profile is $\sigma_{\theta 1}=0.1\,a=42.5~\mathrm{nm}$.  Increasing the width of the phase profile to the order of the soliton healing length will cause the soliton to drift to the left, as we have observed in simulations.

In future studies we would like to consider the continuum limit as suggested by Fleischhauer~\cite{schmidt2007,muth2009}.  Then there is an interesting interplay between holding $\nu U / J$ constant in order to keep a consistent ratio of nonlinearity to dispersion in the mean-field theory, while at the same time holding $U / \nu J$ constant in order to reach the continuum.  This would allow us to make explicit experimental predictions for the Heidelberg and Hamburg groups; at present our simulations remain in the weakly discretized limit and cannot be clearly and rigorously extrapolated to recent experiments.  On the other hand, our predictions are relevant for the specific context we have considered, namely, a 1D lattice in the strongly quantum limit of unity filling or less.
Such systems have already been achieved in ground-breaking experiments~\cite{Esslinger04_PRL_92_130403, Fertig05_PRL_94_120403};
in this context solitons would be engineered in an array of 1D tubes.
Then one runs $\mathcal{O}(100^2)$ dark soliton experiments simultaneously.  Such high statistics can be useful in extracting $g^{(2)}$ from noise correlations~\cite{altman2004}.

We acknowledge use of the time-evolving block decimation algorithm from the TEBD open source project~\cite{tebdOpenSource} and thank Michael Wall and James Williams.  This work was supported by the National Science Foundation under Grant PHY-0547845 as part of the NSF CAREER program (LDC and RVM), and under Physics Frontiers Center Grant PHY-0822671 (CWC).  ID acknowledges support from a Grant-in-Aid from JSPS.  RVM acknowledges fellowship support from Microsoft Station Q and the SURF program at NIST.  We thank Michael Fleischhauer and Masahito Ueda for discussions.

\textit{Note added in press: Three preprints appeared around the time of press on dark soliton quantum dynamics in various regimes~\cite{gangardt,damski,martin}.}


\begin{thebibliography}{10}

\bibitem{Kivshar98_PhysRep_298_81}
Y.~S. Kivshar and B. Luther-Davis, Phys. Rep. {\bf 298},  81  (1998).

\bibitem{lamb1976}
G.~L. Lamb, Phys. Rev. Lett. {\bf 37},  235  (1976).

\bibitem{reinhardt1997}
W.~P. Reinhardt and C.~W. Clark, J. Phys. B: At. Mol. Opt. Phys. {\bf 30},
  L785  (1997).

\bibitem{kevrekidisPG2008}
{\em Emergent Nonlinear Phenomena in Bose-Einstein Condensates}, edited by
  P.~G. Kevrekidis, D.~J. Frantzeskakis, and R. Carretero-Gonz\'alez
  (Springer-Verlag, Berlin, 2008).

\bibitem{Burger99_PRL_83_5198}
S. Burger, K. Bongs, S. Dettmer, W. Ertmer, K. Sengstock, A. Sanpera, G.~V.
  Shlyapnikov, and M. Lewenstein, Phys. Rev. Lett. {\bf 83},  5198  (1999).

\bibitem{Denschlag00_Science_287_97}
J. Denschlag, J.~E. Simsarian, D.~L. Feder, C.~W. Clark, L.~A. Collins, J.
  Cubizolles, L. Deng, E.~W. Hagley, K. Helmerson, W.~P. Reinhardt, S.~L.
  Rolston, B.~I. Schneider, and W.~D. Phillips, Science {\bf 287},  97  (2000).

\bibitem{Sengstock08_NaturePhys_4_496}
C. Becker, S. Stellmer, P. Soltan-Panahi, S. D\"{o}rscher, M. Baumert, E.-M.
  Richter, J. Kronj\"{a}ger, K. Bongs, and K. Sengstock, Nature Phys. {\bf 4},
  496  (2008).

\bibitem{Sengstock08_PRL_101_120406}
S. Stellmer, C. Becker, P. Soltan-Panahi, E.-M. Richter, S. D\"{o}rscher, M.
  Baumert, J. Kronj\"{a}ger, K. Bongs, and K. Sengstock, Phys. Rev. Lett. {\bf
  101},  120406  (2008).

\bibitem{Oberthaler08_PRL_101_130401}
A. Weller, J.~P. Ronzheimer, C. Gross, J. Esteve, M.~K. Oberthaler, D.~J.
  Frantzeskakis, G. Theocharis, and P.~G. Kevrekidis, Phys. Rev. Lett. {\bf
  101},  130401  (2008).

\bibitem{anderson2001}
B.~P. Anderson, P.~C. Haljan, C.~A. Regal, D.~L. Feder, L.~A. Collins, C.~W.
  Clark, and E.~A. Cornell, Phys. Rev. Lett. {\bf 86},  2926  (2001).

\bibitem{ginsberg2005}
N.~S. Ginsberg, J. Brand, and L.~V. Hau, Phys. Rev. Lett. {\bf 94},  040403
  (2005).

\bibitem{Jackson07_PRA_75_051601}
B. Jackson, N.~P. Proukakis, and C.~F. Barenghi, Phys. Rev. A {\bf 75},  051601(R)
   (2007).

\bibitem{Dalfovo99_RevModPhys_71_463}
F. Dalfovo, S. Giorgini, L.~P. Pitaevskii, and S. Stringari, Rev. Mod. Phys.
  {\bf 71},  463  (1999).

\bibitem{Dziarmaga03_JPhysB_36_1217}
J. Dziarmaga, Z.~P. Karkuszewski, and K. Sacha, J. Phys. B: At. Mol. Opt. Phys.
  {\bf 36},  1217  (2003).

\bibitem{Dziarmaga06_JPhysB_39_57}
J. Dziarmaga and K. Sacha, J. Phys. B: At. Mol. Opt. Phys. {\bf 39},  57
  (2006).

\bibitem{Dziarmaga04_PRA_70_063616}
J. Dziarmaga, Phys. Rev. A {\bf 70},  063616  (2004).

\bibitem{Mishmash09_short}
R.~V. Mishmash and L.~D. Carr, Phys. Rev. Lett. {\bf 103}, 140403 (2009).

\bibitem{Vidal04_PRL_93_040502}
G. Vidal, Phys. Rev. Lett. {\bf 93},  040502  (2004).

\bibitem{Lewenstein07_AdvPhys_56_243}
M. Lewenstein, A. Sanpera, V. Ahufinger, B. Damski, A. Sen(De), and U. Sen,
  Adv. Phys. {\bf 56},  243  (2007).

\bibitem{Bloch08_RevModPhys_80_885}
I. Bloch, J. Dalibard, and W. Zwerger, Rev. Mod. Phys. {\bf 80},  885  (2008).

\bibitem{Fisher89_PRB_40_546}
M.~P.~A. Fisher, P.~B. Weichman, G. Grinstein, and D.~S. Fisher, Phys. Rev. B
  {\bf 40},  546  (1989).

\bibitem{Carr08_PRL_100_060401}
R. Kanamoto, L.~D. Carr, and M. Ueda, Phys. Rev. Lett. {\bf 100},  060401
  (2008).

\bibitem{Carr09}
R. Kanamoto, L.~D. Carr, and M. Ueda, Phys. Rev. A {\bf 79},  063616
  (2009).

\bibitem{Carr09b}
R. Kanamoto, L.~D. Carr, and M. Ueda, Phys. Rev. A under review, e-print arXiv:0910.2805 (2009).

\bibitem{Greiner02_Nature_415_39}
M. Greiner, O. Mandel, T. Esslinger, T.~W. H\"{a}nsch, and I. Bloch, Nature
  {\bf 415},  39  (2002).

\bibitem{Jaksch98_PRL_81_3108}
D. Jaksch, C. Bruder, J.~I. Cirac, C.~W. Gardiner, and P. Zoller, Phys. Rev.
  Lett. {\bf 81},  3108  (1998).

\bibitem{Tsuzuki70_JLowTempPhys_4_441}
T. Tsuzuki, J. Low Temp. Phys. {\bf 4},  441  (1971).

\bibitem{ZakharovShabat71}
V.~E. Zakharov and A.~B. Shabat, Zh. Eksp. Teor. Fiz. {\bf 61},  118  (1971),
  [\emph{Sov. Phys. JETP}, 34:62--69, 1972].

\bibitem{ZakharovShabat73}
V.~E. Zakharov and A.~B. Shabat, Zh. Eksp. Teor. Fiz. {\bf 64},  1627  (1973),
  [\emph{Sov. Phys. JETP}, 37:823--828, 1973].

\bibitem{Strecker02_Nature_417_150}
K.~E. Strecker, G.~B. Partridge, A.~G. Truscott, and R.~H. Hulet, Nature {\bf
  417},  150  (2002).

\bibitem{Salomon02_Science_296_1290}
L. Khaykovich, F. Schreck, G. Ferrari, T. Bourdel, J. Cubizolles, L.~D. Carr,
  Y. Castin, and C. Salomon, Science {\bf 296},  1290  (2002).

\bibitem{Bronski01_PRL_89_1402}
J.~C. Bronski, L.~D. Carr, B. Deconinck, and J.~N. Kutz, Phys. Rev. Lett. {\bf
  86},  1402  (2001).

\bibitem{Louis03_PRA_67_013602}
P.~J.~Y. Louis, E.~A. Ostrovskaya, C.~M. Savage, and Y.~S. Kivshar, Phys. Rev.
  A {\bf 67},  013602  (2003).

\bibitem{Efremidis03_PRA_67_063608}
N.~K. Efremidis and D.~N. Christodoulides, Phys. Rev. A {\bf 67},  063608
  (2003).

\bibitem{Trombettoni01_PRL_86_2353}
A. Trombettoni and A. Smerzi, Phys. Rev. Lett. {\bf 86},  2353  (2001).

\bibitem{Kivshar94_PRE_50_5020}
Y.~S. Kivshar, W. Kr\'{o}likowski, and O.~A. Chubykalo, Phys. Rev. E {\bf 50},
  5020  (1994).

\bibitem{Johansson99_PRL_82_85}
M. Johansson and Y.~S. Kivshar, Phys. Rev. Lett. {\bf 82},  85  (1999).

\bibitem{Kevrekidis03_PRA_68_035602}
P.~G. Kevrekidis, R. Carretero-Gonz\'{a}lez, G. Theocharis, D.~J.
  Frantzeskakis, and B.~A. Malomed, Phys. Rev. A {\bf 68},  035602  (2003).

\bibitem{Eiermann04_PRL_92_230401}
B. Eiermann, T. Anker, M. Albiez, M. Taglieber, P. Treutlein, K.~P. Marzlin, and M.~K.
  Oberthaler, Phys. Rev. Lett. {\bf 92},  230401  (2004).

\bibitem{Amico98_PRL_80_2189}
L. Amico and V. Penna, Phys. Rev. Lett. {\bf 80},  2189  (1998).

\bibitem{Dziarmaga02_PRA_66_043615}
J. Dziarmaga, Z.~P. Karkuszewski, and K. Sacha, Phys. Rev. A {\bf 66},  043615
  (2002).

\bibitem{Dziarmaga02_PRA_66_043620}
J. Dziarmaga and K. Sacha, Phys. Rev. A {\bf 66},  043620  (2002).

\bibitem{Law03_PRA_68_015602}
C.~K. Law, Phys. Rev. A {\bf 68},  015602  (2003).

\bibitem{Scott94_PhysD_78_194}
A.~C. Scott, J.~C. Eilbeck, and H. Gilhoj, Phys. D {\bf 78},  194  (1994).

\bibitem{Lai89_PRA_40_844}
Y. Lai and H.~A. Haus, Phys. Rev. A {\bf 40},  844  (1989).

\bibitem{Mishmash08_IMACS}
R.~V. Mishmash and L.~D. Carr, Math. Comp. Sim., in press, doi:10.1016/j.matcom.2009.08.025, e-print arXiv:0810.2593 (2008).

\bibitem{Vidal03_PRL_91_147902}
G. Vidal, Phys. Rev. Lett. {\bf 91},  147902  (2003).

\bibitem{Mishmash08_MSThesis}
R.~V. Mishmash, Master's thesis, Colorado School of Mines, 2008.

\bibitem{Pegg89_PRA_39_1665}
D.~T. Pegg and S.~M. Barnett, Phys. Rev. A {\bf 39},  1665  (1989).

\bibitem{Einstein35_EPRpaper}
A. Einstein, B. Podolsky, and N. Rosen, Phys. Rev. {\bf 47},  777  (1935).

\bibitem{Brennen03_QuantInfoAndComp_3_619}
G.~K. Brennen, Quant. Inf. and Comp. {\bf 3},  619  (2003).

\bibitem{Barnum03_PRA_68032308}
H. Barnum, E. Knill, G. Ortiz, and L. Viola, Phys. Rev. A {\bf 68},  032308
  (2003).

\bibitem{Barnum04_PRL_92_107902}
H. Barnum, E. Knill, G. Ortiz, R. Somma, and L. Viola, Phys. Rev. Lett. {\bf
  92},  107902  (2004).

\bibitem{Hu06_JMathPhys_47_023502}
X. Xi and Z. Ye, J. Math. Phys. {\bf 47},  023502  (2006).

\bibitem{Tsallis88_JStatPhys_52_479}
C. Tsallis, J. Stat. Phys. {\bf 52},  479  (1988).

\bibitem{Carr01_PRA_63_051601}
L.~D. Carr, J. Brand, S. Burger, and A. Sanpera, Phys. Rev. A {\bf 63},  051601(R)
   (2001).

\bibitem{Daley05_PhDThesis}
A.~J. Daley, Ph.D. thesis, Leopold-Franzens-Universit{\"a}t Innsbruck, 2005.

\bibitem{Kollath05_PRA_71_053606}
C. Kollath, U. Schollw\"ock, J. von Delft, and W. Zwerger, Phys. Rev. A {\bf
  71},  053606  (2005).

\bibitem{Danshita09_PRA_79_043601}
I. Danshita and P. Naidon, Phys. Rev. A {\bf 79},  043601  (2009).

\bibitem{carr2002a}
S. Burger, L.~D. Carr, P. \"Ohberg, K. Sengstock, and A. Sanpera, Phys. Rev. A
  {\bf 65},  043611  (2002).

\bibitem{Busch00_PRL_84_2298}
T. Busch and J.~R. Anglin, Phys. Rev. Lett. {\bf 84},  2298  (2000).

\bibitem{Raghavan99_PRA_59_620}
S. Raghavan, A. Smerzi, S. Fantoni, and S.~R. Shenoy, Phys. Rev. A {\bf 59},
  620  (1999).

\bibitem{Ichihara08_PRA_78_063604}
R. Ichihara, I. Danshita, and T. Nikuni, Phys. Rev. A {\bf 78},  063604
  (2008).

\bibitem{schmidt2007}
B. Schmidt and M. Fleischhauer, Phys. Rev. A {\bf 75},  021601(R)  (2007).

\bibitem{muth2009}
D. Muth, B. Schmidt, and M. Fleischhauer, e-print arXiv:0910.1749 (2009).

\bibitem{Esslinger04_PRL_92_130403}
T. St\"oferle, H. Moritz, C. Schori, M. K\"ohl, and T. Esslinger, Phys. Rev.
  Lett. {\bf 92},  130403  (2004).

\bibitem{Fertig05_PRL_94_120403}
C.~D. Fertig, K.~M. O'Hara, J.~H. Huckans, S.~L. Rolston, W.~D. Phillips, and
  J.~V. Porto, Phys. Rev. Lett. {\bf 94},  120403  (2005).

\bibitem{altman2004}
E. Altman, E. Demler, and M.~D. Lukin, Phys. Rev. A {\bf 70},  013603  (2004).

\bibitem{tebdOpenSource}
http://physics.mines.edu/downloads/software/tebd/.

\bibitem{gangardt}
D. M. Gangardt and A. Kamenev, e-print arXiv:0908.4513 (2009).

\bibitem{damski}
B. Damski and W. H. Zurek, e-print arXiv:0909.0761 (2009).

\bibitem{martin}
A. D. Martin and J. Ruostekoski, e-print arXiv:0909.2621 (2009).
\end{thebibliography}
\end{document}